\documentclass[aps, pra, superscriptaddress, 10pt, twocolumn]{revtex4-2} 

\usepackage{amsmath}
\usepackage{amsthm}
\usepackage{bm}
\usepackage{graphicx}
\usepackage{amsfonts}
\usepackage{amssymb}
\usepackage{braket}
\usepackage{textcomp}
\usepackage[toc]{appendix}
\usepackage{cancel}
\usepackage{xcolor}
\usepackage{pifont}
\usepackage[normalem]{ulem}
\usepackage[colorlinks=true,linkcolor=blue,urlcolor=blue,citecolor=blue,pdfusetitle]{hyperref}
\usepackage[all]{hypcap} 

\newcommand{\eq}[1]{\text{Eq.}~\eqref{#1}}

\newcommand{\subf}[2]{\text{Fig.}~\ref{#1}\textcolor{blue}{#2}}

\newcommand{\sect}[1]{\text{Sec.}~\ref{#1}}

\newcommand{\tr}[1]{\text{Tr}\left[#1\right]}

\newcommand{\iden}{\mathbb{I}}

\newcommand{\mD}{\mathcal{D}}
\newcommand{\mJ}{\mathcal{J}}
\newcommand{\mL}{\mathcal{L}}
\newcommand{\mQ}{\mathcal{Q}}
\newcommand{\mU}{\mathcal{U}}

\newcommand{\mauroc}[1]{\textcolor{blue}{}}

\newcommand{\matteoc}[1]{\textcolor{purple}{}}

\newcommand\D{\operatorname{d}\!}

\begin{document}

\title{Non-equilibrium quantum thermodynamics of a particle trapped in a controllable time-varying potential}

\author{Qiongyuan Wu}
\thanks{Email: qwu03@qub.ac.uk}
\affiliation{Centre for Theoretical Atomic, Molecular and Optical Physics, School of Mathematics and Physics, Queen’s University Belfast, BT7 1NN Belfast, United Kingdom}
\author{Luca Mancino}
\affiliation{Centre for Theoretical Atomic, Molecular and Optical Physics, School of Mathematics and Physics, Queen’s University Belfast, BT7 1NN Belfast, United Kingdom}
\author{Matteo Carlesso}
\affiliation{Centre for Theoretical Atomic, Molecular and Optical Physics, School of Mathematics and Physics, Queen’s University Belfast, BT7 1NN Belfast, United Kingdom}
\author{Mario A. Ciampini}
\affiliation{Vienna Center for Quantum Science and Technology (VCQ), Faculty of Physics, University of Vienna, 1090 Vienna, Austria}
\author{Lorenzo Magrini}
\affiliation{Vienna Center for Quantum Science and Technology (VCQ), Faculty of Physics, University of Vienna, 1090 Vienna, Austria}

\author{Nikolai Kiesel}
\affiliation{Vienna Center for Quantum Science and Technology (VCQ), Faculty of Physics, University of Vienna, 1090 Vienna, Austria}
\author{Mauro Paternostro}
\affiliation{Centre for Theoretical Atomic, Molecular and Optical Physics, School of Mathematics and Physics, Queen’s University Belfast, BT7 1NN Belfast, United Kingdom}

\date{\today}

\begin{abstract}
Many advanced quantum techniques feature non-Gaussian dynamics, and the ability to manipulate the system in that domain is the next-stage in many experiments. One example of meaningful non-Gaussian dynamics is that of a double-well potential. Here we study the dynamics of a levitated nanoparticle undergoing the transition from an harmonic potential to a double-well in a realistic setting, subjecting to both thermalisation and localisation.  We characterise the dynamics of the nanoparticle from a thermodynamic point-of-view, investigating the dynamics with the Wehrl entropy production and its rates. Furthermore, we investigate coupling regimes where the the quantum effect and thermal effect are of the same magnitude, and look at suitable squeezing of the initial state that provides the maximum coherence. The effects and the competitions of the unitary and the dissipative parts onto the system are demonstrated. 
We quantify the requirements to relate
our results to a bonafide experiment with the presence of the environment, and discuss the experimental interpretations of our results in the end.

\end{abstract}

\maketitle


\section{Introduction}
The potential for enhanced performances above and beyond the possibilities offered by classical devices underpins the development of quantum technologies for applications, in information and communication technology, sensing, and computation
\cite{nielsen_chuang_2010, Shor_1997, Grover_1997, 
Giovannetti_2004, Giovannetti_2006, Giovannetti_2011}.
Recently, it has been realized that quantum laws could also be exploited to manage more efficiently the energetics of nanoscale technologies, thus contributing to the emergence of a framework for the thermodynamics of quantum processes~\cite{Kosloff_2013, Campo_2014, Rossnagel_2014, Rossnagel_2016, Deffner_2019, Vinjanampathy_2016, Goold_2016}. However, such advantages do not come easily due to unavoidable necessity to protect the mechanisms responsible for the quantum process being studied from detrimental environmental effects~
\cite{Zeh_1970, Zurek_2003, Schlosshauer_2005, Schlosshauer_2019}. 
Such deleterious influences have a tendency to scale with the size of the system being used, and become -- in principle -- prohibitively severe in the mesoscopic and macroscopic regimes. 

Fortunately, impressive progress on the experimental control of quantum systems and dynamics have been achieved in the course of the past twenty years. This has allowed the design and implementation of effective strategies for the control of mesoscopic systems such as cold atoms \cite{UltraColdReview_2020}, large arrays of superconducting systems \cite{SuperconductingReview_2019, SuperconductingReview_2021}, and (electro-/opto-/magneto-)mechanical structures \cite{CavityReview_2014, LevitoReview_2021}. 
Levito-dynamics \cite{LevitoReview_2021}, i.e., the levitation of nano- and micro-objects in vacuum, holds the potential  to become a key experimental platform for the demonstration of quantum features at the mesoscopic scale. It enables dynamical control over nearly arbitrary potentials as well as high degrees of isolation from environmental influences. Today, experiments based on optical-tweezer technology \cite{NanoThermoReview_2018} have allowed near Heisenberg-limited position readout, real-time control of motion and preparation in the ground state of motion\cite{Delic_2020, Magrini_2021, Tebbenjohanns_2021}. Also, dynamical shaping of an optical potential for a levitated nanoparticle has been demonstrated, with the scope to implement a logically irreversible transformation~\cite{Ciampini_2021}. These developments pave the way to the exploration of non-equilibrium phenomena in open mesoscopic systems and thus the consolidation of a framework of controllable quantum thermodynamics of large-scale systems~\cite{Brunelli_2018}.

In this work, we make significant theoretical steps towards the characterization of such framework by addressing thermodynamic irreversibility stemming from the non-equilibrium process involving a levitated nanoparticle -- subjected to both thermalisation and decoherence mechanisms -- in a potential landscape that is changing from a quadratic to a double-well configuration, similar to the experimental situation in~\cite{Ciampini_2021}. We focus our attention on the quantification of irreversible entropy production~\cite{Landi_2020}, a key quantity that characterizes logical and thermodynamic irreversibility, constraints the performances of quantum and classical engines, and appears to be strongly related to the occurrence of non-equilibrium quantum critical phenomena~\cite{Fusco2014,Landi_2020}. Our investigation sets the methodological toolbox for the successful simulation of non-equilibrium processes subjected to real-time potential-shaping transformation, as envisioned in particular for levitated systems. It also establishes the context for the quantification of potential thermodynamics-based limitations to the efficiency of quantum memories.

The remainder of this paper is organized as follows. In \sect{sec:modelandmethod} we introduce the model and introduce a suitable angular-momentum algebra that allows for an agile simulation of the dynamics of the system. We illustrate such capabilities through significant numerical examples.
In \sect{sec:wehrlentropy}, we introduce the thermodynamic quantities of interest, namely the Wehrl entropy, its production rate and the entropy flux rate~\cite{Landi_2020}. These form a toolbox of figures of merit for the characterization of thermodynamic irreversibility stemming from the non-equilibrium process undergone by the system, and the energy-exchanging mechanisms with the environment resulting from its open-system dynamics. In \sect{sec:numericalsimulation}, we focus on a quantitative characterization of such quantities for a system 
 undergoing temporal shaping of its trapping potential from  harmonic to double-well shape. An experimental perspective is given in \sect{sec:experimentperspective}, while we draw our conclusions in \sect{sec:conclusions}.

\section{The Model}\label{sec:modelandmethod}

We consider a 
mechanical system under the influences of the surrounding environment. 
The dynamics of the system can be accounted by the following master equation for the density matrix $\rho$,
\begin{equation}\label{equ:systemdynamics}
\dot{\rho} = -\frac{i}{\hbar} [H_\text{s}, \rho] + D_\text{th}[\rho] + D_\text{lc}[\rho].
\end{equation}
The second term in Eq.~\eqref{equ:systemdynamics} is the thermalisation dissipator $D_\text{th}[\rho]$, which describes the effect of heat exchange between system and environment and would eventually lead to thermal equilibrium at the environmental temperature.
Such term takes the form of the Lindblad super-operator \cite{Kosloff_2013,Deffner_2019} 
\begin{equation}\label{equ:thermaldissipator}
\begin{split}
   D_\text{th}[\rho] = &\gamma\left[(\bar{n}+1) {\cal L}_{a}(\rho)
    +  \bar{n} {\cal L}_{a^\dag}(\rho)\right]
\end{split}
\end{equation}
with ${\cal L}_O(\rho)=O \rho O^\dagger - \lbrace O^\dagger O , \rho \rbrace/2$ for any operator $O$. Here, $a$ and $a^\dag$ are bosonic annihilation and creation operators of a harmonic oscillator of frequency $\omega$ that will work as a basis of our problem, $\gamma$ is the coupling strength between system and its thermal environment, which has a mean number of excitations $\bar{n} = (e^{\beta \hbar\omega} -1)^{-1}$ with $\beta = 1/k_\text{\tiny B}T$ being the inverse temperature of the environment. 
The last term in Eq.~\eqref{equ:systemdynamics} quantifies the decoherence effects of the collision of the residual gas in the vacuum, and it can be described through a localisation term \cite{Joos_1985, Kiefer_1999, Schlosshauer_2019}, which reads 
\begin{equation}\label{equ:localisationdissipator}
D_\text{lc}[\rho] = - \Lambda[x,[x,\rho]],
\end{equation}
where $\Lambda$ is the coupling strength and $x=\sqrt{\hbar/2m \omega}(a^\dag+a)$ is the position operator. 
Finally, $H_\text{s}(t)$ in Eq.~\eqref{equ:systemdynamics} denotes the system Hamiltonian, whose time dependence can be used to dynamically modify the potential. In particular, we focus on the following form
\begin{equation}\label{equ:systemhamiltonian2}
H_\text{s} (t) = \frac{p^2}{2m} + \frac{1}{2}m\omega^2 x^2 + H_\text{add}(t),
\end{equation}
where we have introduced the momentum operator $p=i\sqrt{\hbar m \omega/2}(a^\dag-a)$. Eq.~\eqref{equ:systemhamiltonian2} describes a harmonic potential modified by an additional time-dependent term whose form we take as
\begin{equation}\label{equ:doublepotential}
H_\text{add}(t) = - {\cal E}\left(\alpha(t) + \frac{\bar \alpha(t)}{2}\frac{x^2}{W^2}\right) e^{-\frac{x^2}{2W^2}}.
\end{equation}
Here, ${\cal E}$ is a suitably chosen energy scale. 
The protocol embodied by the time-variation of $H_\text{add}(t)$ creates a double-well potential from  one of the form of an inverted Gaussian via a suitable change of $\alpha(t)$ and $\bar \alpha(t)$. Here, we define a protocol that linearly switches the potential in time from $0$ to $\tau$, such that  $\bar \alpha(t) = 1 - \alpha(t)$, and $\alpha(t) = 1 - t/\tau$, with $\tau$ the characteristic time of the protocol, and we fix ${\cal E} =10\hbar\omega$ so that the ground state energy is below the central peak of the final double-well potential. The initial potential (at $t=0$) well approximates the low energy-sector of the harmonic oscillator.
The total potential is shown in \subf{fig:doublewellpotentialplot}{a} and we refer to \sect{sec:experimentperspective} for experimental considerations on the plausiblity of the choices made here.

\begin{figure*}
    \centering
    \begin{minipage}[b]{0.32\textwidth}
        \textbf{(a)}
    \end{minipage}%
    \hfill%
    \begin{minipage}[b]{0.32\textwidth}
        \textbf{(b)}
    \end{minipage}
    \hfill%
    \begin{minipage}[b]{0.32\textwidth}
        \textbf{(c)}
    \end{minipage}\\[1ex]
    \begin{minipage}[t]{0.32\textwidth}
        \includegraphics[width=\textwidth]{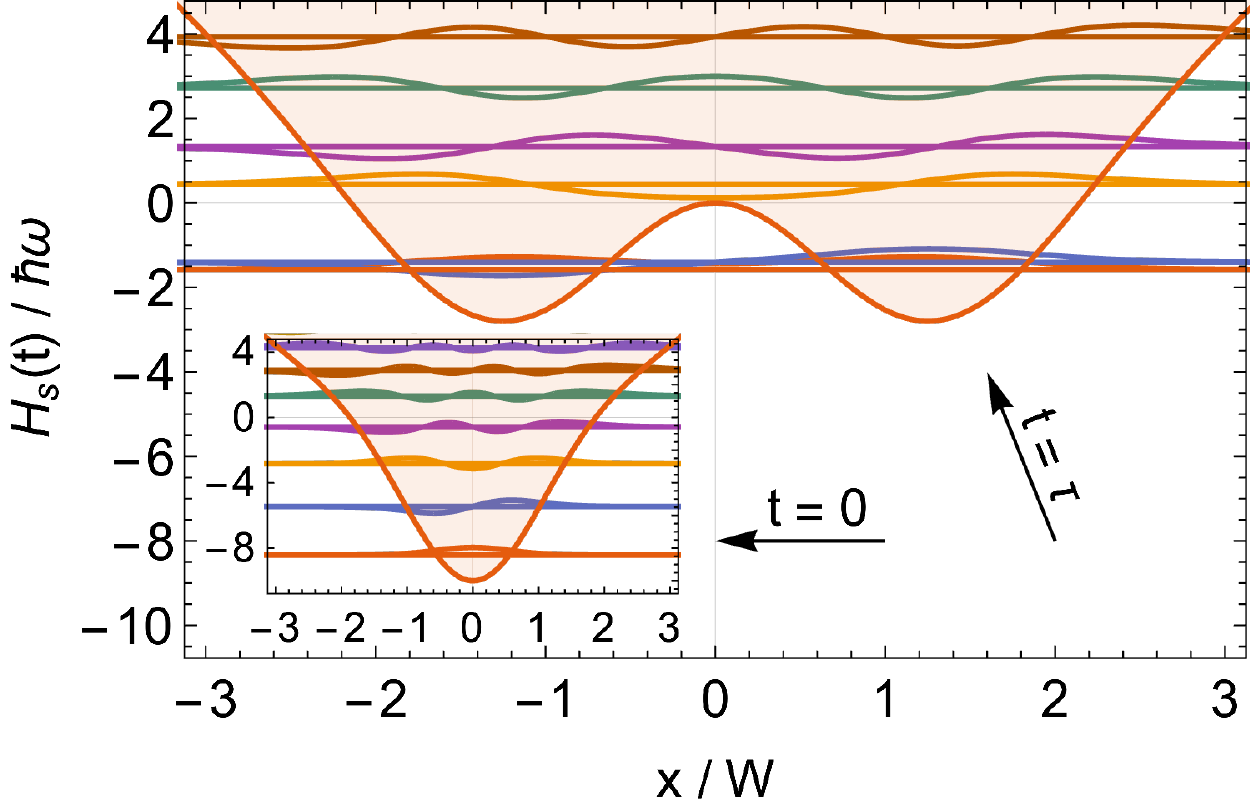}
    \end{minipage}%
    \hfill%
    \begin{minipage}[t]{0.32\textwidth}
        \includegraphics[width=\textwidth]{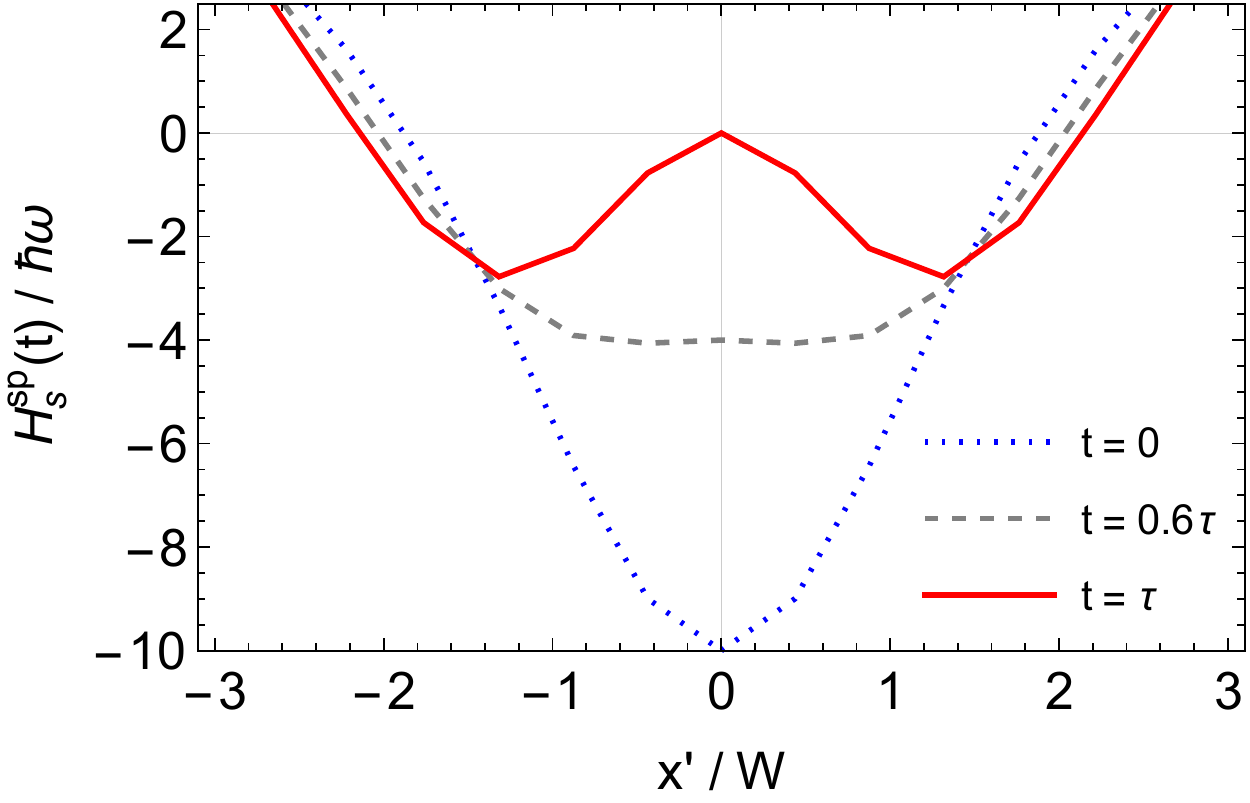}
    \end{minipage}
    \hfill%
    \begin{minipage}[t]{0.32\textwidth}
        \includegraphics[width=\textwidth]{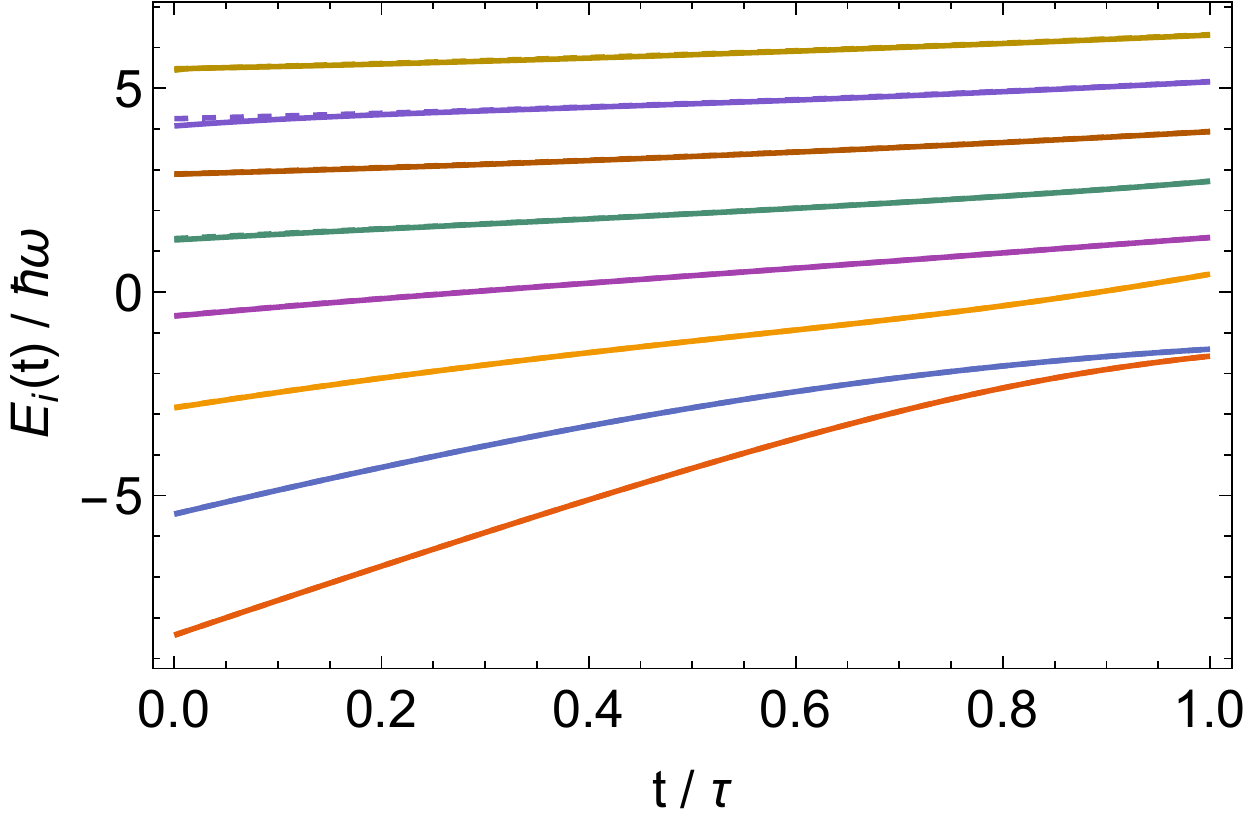}
    \end{minipage}
	\caption{Time-varying potential and the instantaneous eigenenergies of the continuous and discretized Hamiltonians. 
	Panel~{\bf (a)} shows the potential and eigenenergies change in continuous position picture. The potential starts from an inverted Gaussian form at $t=0$, then gradually deforms into a double well until $t=\tau$.
	Panel~{\bf (b)} shows the discretized potential change by Eq.~\eqref{equ:qohamiltonianinqsc} and \eq{equ:doublepotentiainqsc}, with dimension $N=25$ (or $j=12$) and expansion order $\kappa=15$. It captures the same dynamics as the continuous situation. 
	Panel~{\bf (c)} compares the first 8 instantaneous eigenenergies of the continuous and discretized Hamiltoinians, which are respectively shown with dashed and solid lines.}
	\label{fig:doublewellpotentialplot}
\end{figure*}

%
In order to analyze the dynamics of the system as the shape of the trapping potential varies in time, we resort to a spin-coherent state picture that provides an accurate effective description of the dynamics of the system in the low-energy sector of the Hilbert space. This is done through the Holstein–Primakoff (HP) transformation, shown as follows. The discretization of the quantum oscillator system is helpful for numerical simulations, as well as the assessment of the non-equilibrium thermodynamics of the process, as we explain in Sec.~\ref{sec:wehrlentropy}.


\subsection*{Transformation from harmonic to angular momentum description}
Computing dynamics of a quantum system with a varying potential under the action of an environment is in general difficult due to the infinite dimensions of the Hilbert space. The general non-Gaussian nature of the time-dependent potential prevents the use of techniques typical of Gaussian-state analysis, requiring the consideration of high-order moments of the position and momentum operators for a faithful account of the features of the system. 

In order to bypass these bottleneck and provide an agile and physically intuitive picture of the dynamics at hand and its consequences, we aim to describe the system and the potential in Eqs.~\eqref{equ:systemhamiltonian2} and \eqref{equ:doublepotential} through an effective picture based on the physics of angular momenta. With this aim in mind, we proceed as follows: 1.~we discretize the system, and 2.~we write the discretized system in terms of angular momentum operators. 

\textit{Step~1:} 
Discretization of the Hamiltonian can be achieved through the Holstein Primakoff (HP) transformation \cite{Holstein_1940,Gyamfi_2019,Vogl_2020}, which intorduces an effective bosonic system with a fixed dimension $N$. Such HP bosonic system is built with annihilation and creation operators $b$ and $b^\dag$ such that
\begin{equation}\label{equ:hpoperatros}
    \begin{aligned}
       & b\ket{n}=\sqrt{n}\ket{n-1}\qquad\text{for} \quad 1\leq n\leq N, \\
      &  b^\dagger\ket{n} = \sqrt{n+1}\ket{n+1}\quad\text{for} \quad 0\leq n\leq N-1,
    \end{aligned}
\end{equation}
through which one can define the HP position and momentum operators  $x_\text{\tiny HP}=(b^\dag+b)\sqrt{\hbar/2m \omega}$ and  $p_\text{\tiny HP}=i(b^\dag-b)\sqrt{\hbar m \omega/2}$.
Note that the commutation relation between such operator reads $[b,b^\dag]=\iden-N\ket{N}\bra{N}$, where $\iden$ is the identity matrix. We are now able to discretize the system by recasting Eqs.~\eqref{equ:systemhamiltonian2} and~\eqref{equ:doublepotential} through the HP bosonic operators as $x\to x_\text{\tiny HP}$ and $y\rightarrow y_\text{\tiny HP}$. The mapping well approximates the low energy sector as far as the value of $N$ is sufficiently large. For $N\rightarrow\infty$ the mapping is exact.

\textit{Step~2:} 
Now, we consider a spin-$J$ system with spin operators $\{J_x, J_y, J_z, J^2\}$ (the same approach works for general angular momentum operators), we denote with $j$ and $j_z$ the quantum numbers of $J^2$ and $J_z$, respectively, whose simultaneous eigenstates $\ket{j,j_z}$ will be used as computational basis of our problem. The action of $J_z$ and $J_\pm = J_x \pm i J_y$ is defined as usual through
\begin{equation}
    \begin{split}
        J_z\ket{j,j_z}&=\hbar j_z\ket{j,j_z},\\
        J_+\ket{j,j_z}&=\hbar \sqrt{(j-j_z)(j+j_z+1)}\ket{j,j_z},\\
        J_-\ket{j,j_z}&=\hbar \sqrt{(j+j_z)(j-j_z+1)}\ket{j,j_z}.
    \end{split}
\end{equation}
Moreover, the standard commutation relations are satisfied $[J_i,J_k]=i\hbar\epsilon_{ikm}J_m$, where $\{i,j,k\}=\{x,y,z\}$ and $\epsilon_{ikm}$ is the Levi-Civita symbol. To connect this algebra to that of the discrete system, we fix the value of $j=(N-1)/2$, while $j_z=-j,-j+1,\dots,j-1,j$. Then, the HP transformation imposes the relations between the HP bosonic operators and the spin operators as
\begin{equation}\label{equ:transformJ}
    J_z=\hbar(j-b^\dagger b); J_+= \hbar\sqrt{2j-b^\dagger b}\,b; J_-=\hbar b^\dagger\sqrt{2j-b^\dagger b}.
\end{equation}
We notice that $J_+$ and $J_-$ are respectively proportionally mapped to $b$ and $b^\dagger$.  
To express the HP bosonic operators with spin operators, we perform the Taylor expansion on the non-linear term in \eq{equ:transformJ} up to the order $\kappa$, thus obtaining
\begin{equation}
 \hbar\sqrt{2j-b^\dag b}=M_\kappa +     \mathcal{O}((b^\dag b)^{\kappa}),
\end{equation}
where $M_\kappa$ is a $\kappa$-th order polynomial in $b^\dag b$.
To make an explicit example, with an expansion up to the second order, we have 
\begin{equation}
M_2=\hbar\sqrt{2j}-\frac{\hbar}{2\sqrt{2j}}b^\dag b - \frac{\hbar}{8\sqrt{(2j)^3}} (b^\dag b)^2,
\end{equation}
For any order $\kappa$, $M_\kappa$ is a real and diagonal matrix, whose inverse matrix $M_\kappa^{-1}$ can be calculated. Then, we can approximate
\begin{equation}
    b\simeq M_\kappa^{-1}J_+,\quad\text{and} \quad b^\dag\simeq J_-M_\kappa^{-1},
\end{equation}
and correspondingly the position and momentum operators become 
\begin{equation}
    \begin{split}
            J_{x'}&\simeq\frac{1}{\sqrt{2}}(J_-M_\kappa^{-1} + M_\kappa^{-1}J_+), \\ J_{y'}&\simeq\frac{i}{\sqrt{2}}(J_-M_\kappa^{-1} - M_\kappa^{-1}J_+). 
    \end{split}
\end{equation}
This approximation becomes exact for $\kappa\to\infty$. Finally, Step 2 is performed by the following mapping $x_{hp}\rightarrow J_{x'}$ and $p_{hp}\rightarrow J_{y'}$. 
Thus, we can rewrite the dynamics in terms of the spin operators. In particular, the system Hamiltonian defined in Eq.~\eqref{equ:systemhamiltonian2} and \eq{equ:doublepotential} becomes
\begin{equation}\label{equ:qohamiltonianinqsc}
    H^\text{sp}_\text{s}(t) = \frac{J_{y'}^2}{2m} + \frac{1}{2}m\omega^2 J_{x'}^2 + H^\text{sp}_\text{add}(t)
\end{equation}
with 
\begin{equation}\label{equ:doublepotentiainqsc}
    H^\text{sp}_\text{add}(t) = - {\cal E}\left(\alpha(t)  + \bar \alpha(t)\frac{J_{x'}^2}{2W^2}\right) e^{-\frac{J_{x'}^2}{2W^2}},
\end{equation}
where the label sp stands for spin. Similarly, one can easily recast the two dissipators with the spin ladder operators to
\begin{equation}\label{equ:thermomastereqsc}
D_\text{th}[\rho] = \gamma \left[ (\bar{n}+1) \mL_{J_-}(\rho) + \bar{n}\mL_{J_+}(\rho) \right],
\end{equation}
and
\begin{equation}\label{equ:localmastereqsc}
D_\text{lc}[\rho] = - \Lambda[J_x,[J_x,\rho]].
\end{equation}

As we already pointed out, we are focusing on the low-energy sector, thus the value of $N$ can be rather low. To have an accurate description of the first $8$ energy levels, we can  
take the spin system with $N=25$ levels (namely $j=12$) and the Taylor expansion to the order $\kappa=15$. In \subf{fig:doublewellpotentialplot}{c}, we compare the first $8$ energy levels of the continuous Hamiltonian defined in \eq{equ:systemhamiltonian2} with those of the Hamiltonian in terms of the discrete spin degrees of freedom as constructed in \eq{equ:qohamiltonianinqsc}, which underlines the good approximation of our approach throughout the time of the process.

\section{Wehrl Entropy in terms of Q-representation}\label{sec:wehrlentropy}
To characterise the non-equilibrium dynamics of the system, we focus on the entropy production of the process under study. Here, we consider the Wehrl entropy, which is  qualitatively similar to the von Neumann entropy, although better at capturing non-equilibrium features beyond the Gaussian regime. Generally, the Wehrl entropy provides an upper bound to the von Neumann one, and the two coincide for coherent states. 
The Wehrl entropy can be calculated through spin-coherent states, which are defined as \cite{Santos_2018,Landi_2020,Radcliffe_1971}
\begin{equation}
\ket{\Omega} = e^{- i \phi J_z} e^{-i \theta J_y} \ket{j,j}.
\end{equation}
Here $\ket{j,j}$ is the angular momentum state with largest quantum number of $J_z$, and $\Omega=\{\theta, \phi \}$ is the set of Euler angles identifying the direction along which the coherent state points. Thus, the  Wehrl entropy for a system $N=2j+1$ degrees of freedom, as it is the one under investigation, reads \cite{Wehrl_1979, Brunelli_2018,Santos_2018}
\begin{equation}\label{equ:wehrlentropy}
S_\mQ = -\frac{ N}{4\pi} \int \D\Omega~\mQ(\Omega) \ln \mQ(\Omega),
\end{equation}
where $\D\Omega=\sin(\theta)\,\D\theta\,\D\phi$, while $\mQ(\Omega)$ is the Husimi-Q function associated with the state $\rho$, which is defined as \begin{equation}\label{equ:qfunction}
\mQ(\Omega) = \bra{\Omega}\rho\ket{\Omega}.
\end{equation}
Using the spin-coherent state picture, the key terms in the dynamical equations of motion formulated in the density-matrix representation can be mapped to the phase-spaces representation, such that
\begin{equation}\label{equ:correspondances}
\begin{split}
\left[ J_+, \rho \right] &\rightarrow \mJ_+(\mQ) = e^{i\phi} (\partial_\theta + i \cot \theta \partial_\phi) \mQ, \\
\left[ J_-, \rho \right] &\rightarrow \mJ_-(\mQ) = -e^{-i\phi} (\partial_\theta - i \cot \theta \partial_\phi) \mQ, \\
\left[ J_z, \rho \right] &\rightarrow \mJ_z(\mQ) = -i {\partial}_\phi \mQ, \\
\left[ J_x, \rho \right] &\rightarrow \mJ_x(\mQ) = \frac{1}{2} \left( \mJ_+(\mQ) + \mJ_-(\mQ) \right).
\end{split}
\end{equation}
With such correspondences at hand, we are allowed to access thermodynamic aspects of the system, namely irreversible entropy production and entropy flux rates, as stated in the following.
\eq{equ:systemdynamics} can now be reformulated in terms of the Husimi function as 
\begin{equation}\label{equ:qfunctiondynamics}
    \partial_t \mQ=\bra{\Omega} \dot{\rho} \ket{\Omega} 
    = \mU_\mQ+\mD_{th}(\mQ)+\mD_{lc}(\mQ)
\end{equation}
with the terms related to the non-unitary mechanisms being
 $\mD_i(\mQ)= \bra{\Omega} D_{i}[\rho]\ket{\Omega}$ 
 ($i=\text{th, lc}$), and $\mU_\mQ$  the transformed version of the term $-\tfrac{i}{\hbar}[H_\text{s},\rho]$ inducing the unitary evolution of the system. Correspondingly, one obtains three contributions to the rate of the Wehrl entropy production, which can be identified by merging the time derivative of \eq{equ:wehrlentropy} with \eq{equ:qfunctiondynamics}, thus finding
\begin{equation}\label{equ:decomposewehrlentropy}
    \begin{split}
        \frac{\D S_\mQ}{\D t}&= -\frac{ N}{4\pi} \int \D\Omega~\partial_t\mQ\ln{\mQ}, \\
        &=\frac{\D S_\mU}{\D t}+\frac{\D S_{\mD_\text{th}}}{\D t}+\frac{\D S_{\mD_\text{lc}}}{\D t}.
    \end{split}
\end{equation}
The dissipative Wehrl entropy rates connected to the dissipators can be further decomposed into the corresponding irreversible entropy production rate $\Pi$ and entropy flux rate $\Phi$ as \cite{Santos_2018,Landi_2020}
\begin{equation}
    \frac{\D S_\mD}{\D t} = \Pi - \Phi.
\end{equation}
The second thermodynamic law states an ever-increasing irreversible entropy production rate $\Pi$, while the whole entropy production rate  of the system $\D S_\mD/\D t$ can be negative if one has a sufficiently large positive entropy flux rate $\Phi$. The explicit forms of these two terms for the thermalisation dissipation read \cite{Santos_2018}
\begin{equation}\label{equ:thermoentropyfluxrate}
\Phi^\text{th}=\gamma\frac{j(2j+1)}{4\pi} \int \D\Omega \sin\theta \left( \frac{2j \mQ \sin\theta}{(2\overline{n}+1)-\cos\theta} - \partial_\theta \mQ \right),
\end{equation} 
and 
\begin{equation}\label{equ:thermoentropyproductionrate}
\begin{split}
\Pi^\text{th}&= {\gamma}\frac{(2j+1)}{8\pi} \int \frac{\D \Omega}{\mQ} \left\{ 
|\mJ_z(\mQ)|^2 \frac{\large[(2\overline{n}+1)\cos\theta-1\large]}{\tan\theta\sin\theta}
\right. \\
&\left.
+\frac{[2 j \mQ \sin\theta {+} (\cos\theta{-} (2\overline{n}+1))\partial_\theta \mQ]^2}{(2\overline{n}+1)-\cos\theta} \right\}.
\end{split}
\end{equation}
For the localization process one has only the irreversible entropy production with no entropy flux. Thus, $\frac{\D S_{\mD_\text{lc}}}{\D t} = \Pi^\text{lc}$, where
\begin{equation}\label{equ:localisationentropyproductionrate}
\Pi^\text{lc} =\Lambda\frac{ N}{4\pi}   \int \D \Omega~\frac{|\mJ_x(\mQ)|^2}{\mQ},
\end{equation}
which was derived in Ref.~\cite{Santos_2018} for the localization in $J_z$ basis while  we derived that in $J_x$ basis in Appendix~\ref{apdsec:derivationofiep}. 
Then, the entropy associated to this process, strictly connected to the loss of coherence in $J_x$ basis, is always increasing.

\section{Entropy production for simulated dynamics}\label{sec:numericalsimulation}

This section presents  the main results of our study, which are the numerical simulations of the dynamics of the system and the corresponding analysis of its thermodynamics. Throughout the simulations, we set the parameters $\hbar=\omega=m=k_\text{\tiny B}=1$, and the processing length time $\tau\omega=10$.

The system is subject to the thermalisation and localisation dissipators, and next we show the effects of these dissipators in terms of their thermodynamic aspects, i.e.~the Wehrl entropy, the irreversible entropy production rate and the entropy flux rate. To fully understand their respective actions, we look at the thermodynamics in different coupling regimes. When both the dissipators are considered, the system is driven to different non-equilibrium steady states depending on the relative coupling strengths. 

Before digging into the problem at hand, 
we start the simulations with a study case of a spin system, which is simulated under three different conditions: when it is under the action of thermalization; that of localization; and finally under the combined action of the two dissipators. Then, we move back to the quantum oscillator case, where we instead focus 
on two cases: when the system is isolated, where we highlight the effects ot the dynamical change in the potential, and when the system is subject to the combined action of the two dissipators.


\subsection{Spin-J systems} \label{subsect.spin}

\begin{figure*}[tb!]
	\centering
    \begin{minipage}[b]{0.32\textwidth}
        \textbf{(a)}
    \end{minipage}
    \hfill
    \begin{minipage}[b]{0.32\textwidth}
        \textbf{(b)}
    \end{minipage}
    \hfill
    \begin{minipage}[b]{0.32\textwidth}
        \textbf{(c)}
    \end{minipage}\\[1ex]
     \begin{minipage}[t]{0.32\textwidth}
        \includegraphics[width=\textwidth]{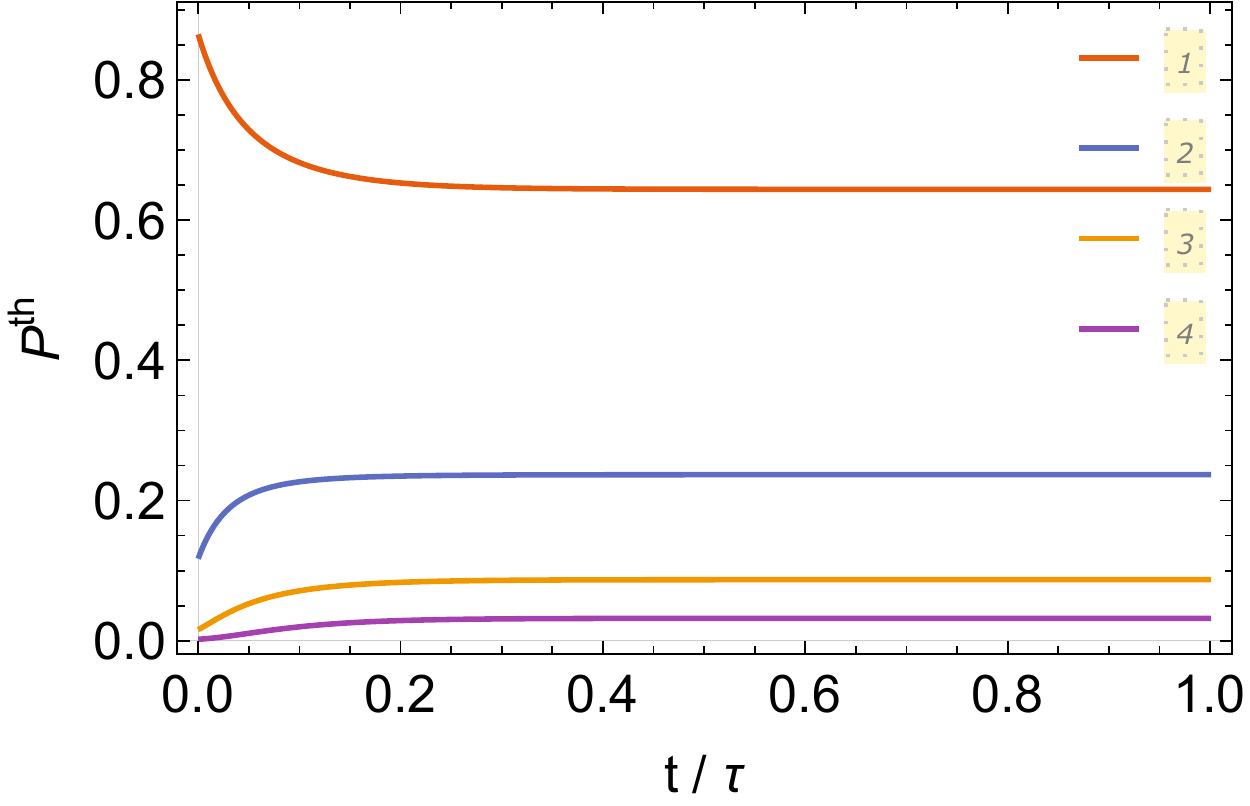}
    \end{minipage}
    \hfill
    \begin{minipage}[t]{0.32\textwidth}
        \includegraphics[width=\textwidth]{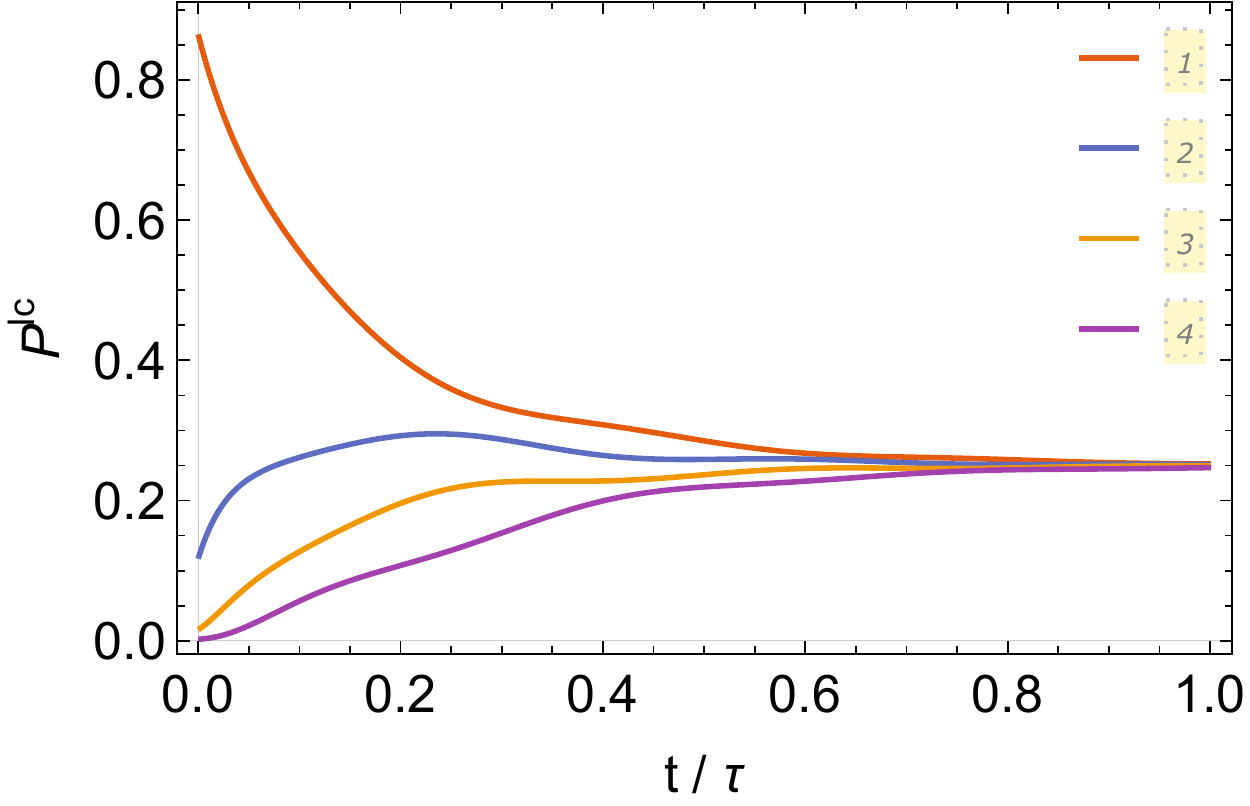}
    \end{minipage}
    \hfill
    \begin{minipage}[t]{0.32\textwidth}
        \includegraphics[width=\textwidth]{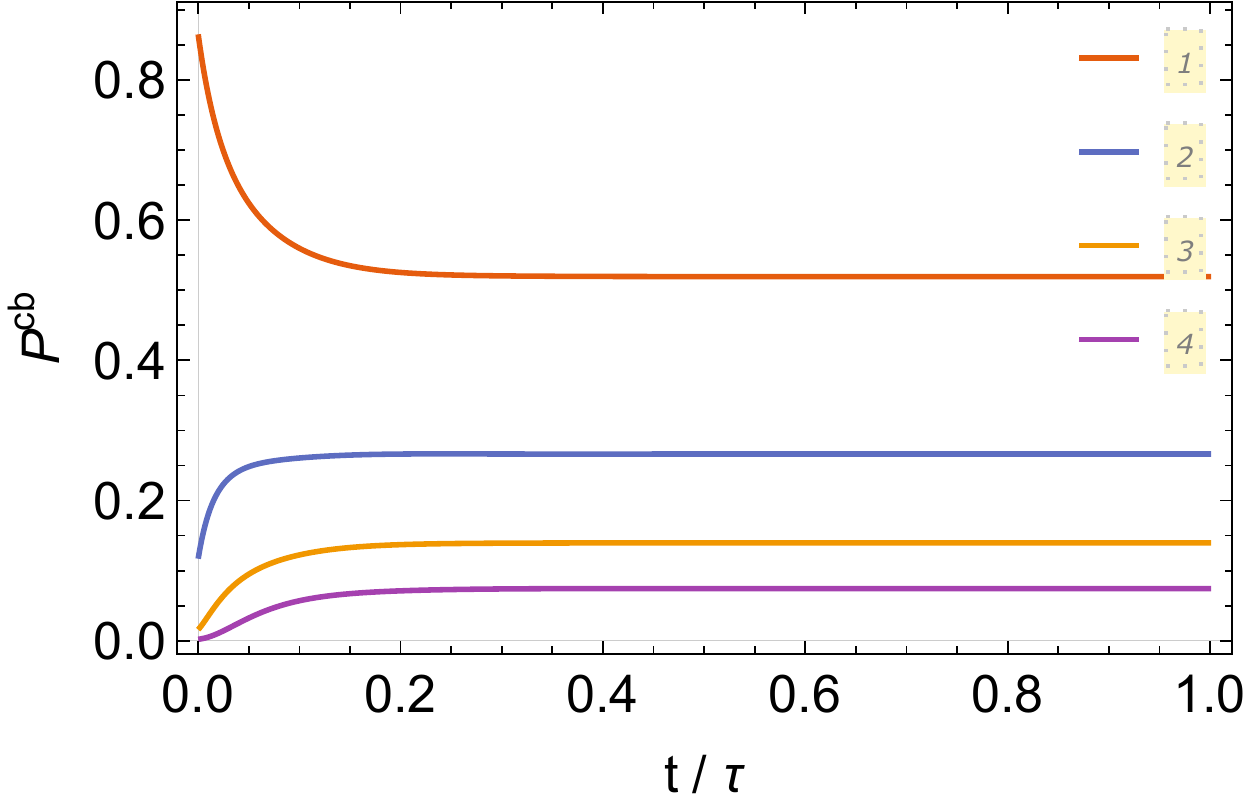}
    \end{minipage}\\[2ex]
 \begin{minipage}[b]{0.32\textwidth}
        \textbf{(d)}
    \end{minipage}%
    \hfill%
    \begin{minipage}[b]{0.32\textwidth}
        \textbf{(e)}
    \end{minipage}
    \hfill%
    \begin{minipage}[b]{0.32\textwidth}
        \textbf{(f)}
    \end{minipage}\\[1ex]
     \begin{minipage}[t]{0.32\textwidth}
        \centering
       \includegraphics[width=\textwidth]{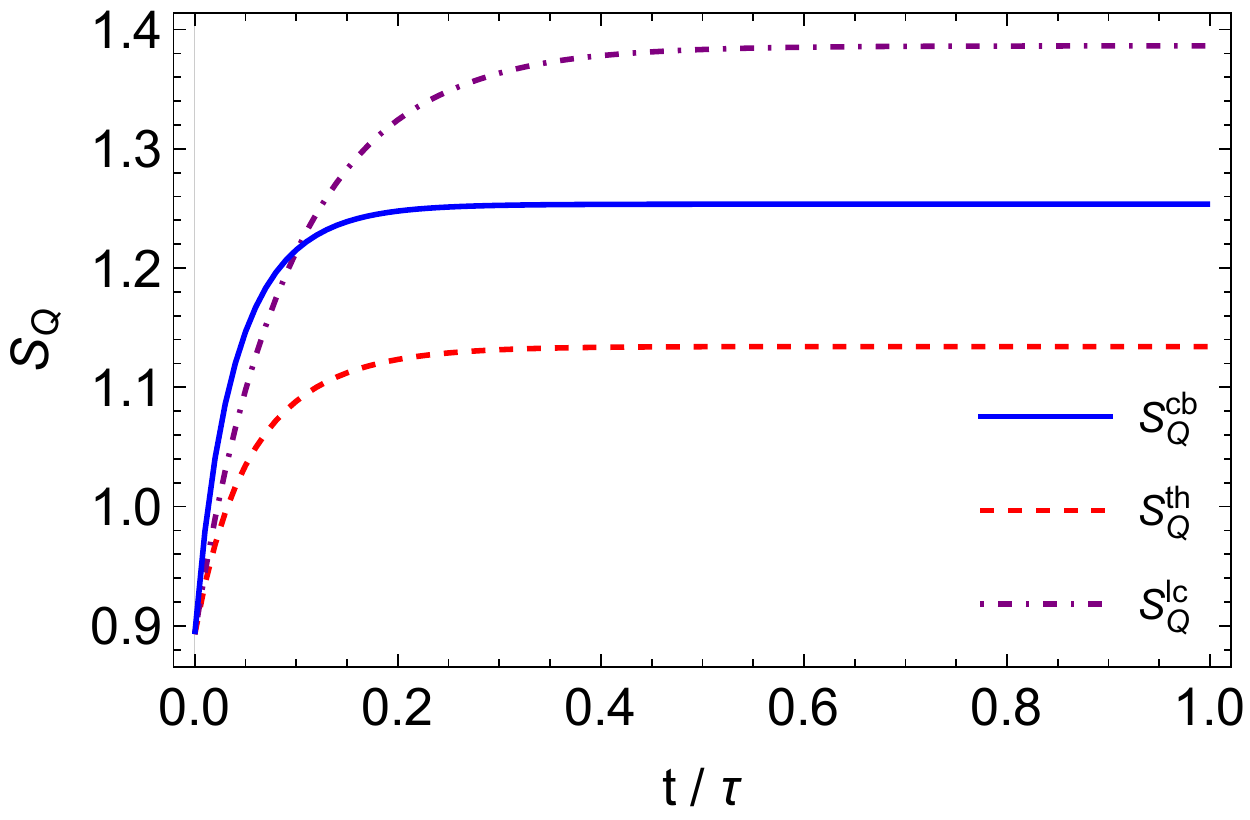}
    \end{minipage}%
    \hfill%
    \begin{minipage}[t]{0.32\textwidth}
    \centering
        \centering
       \includegraphics[width=\textwidth]{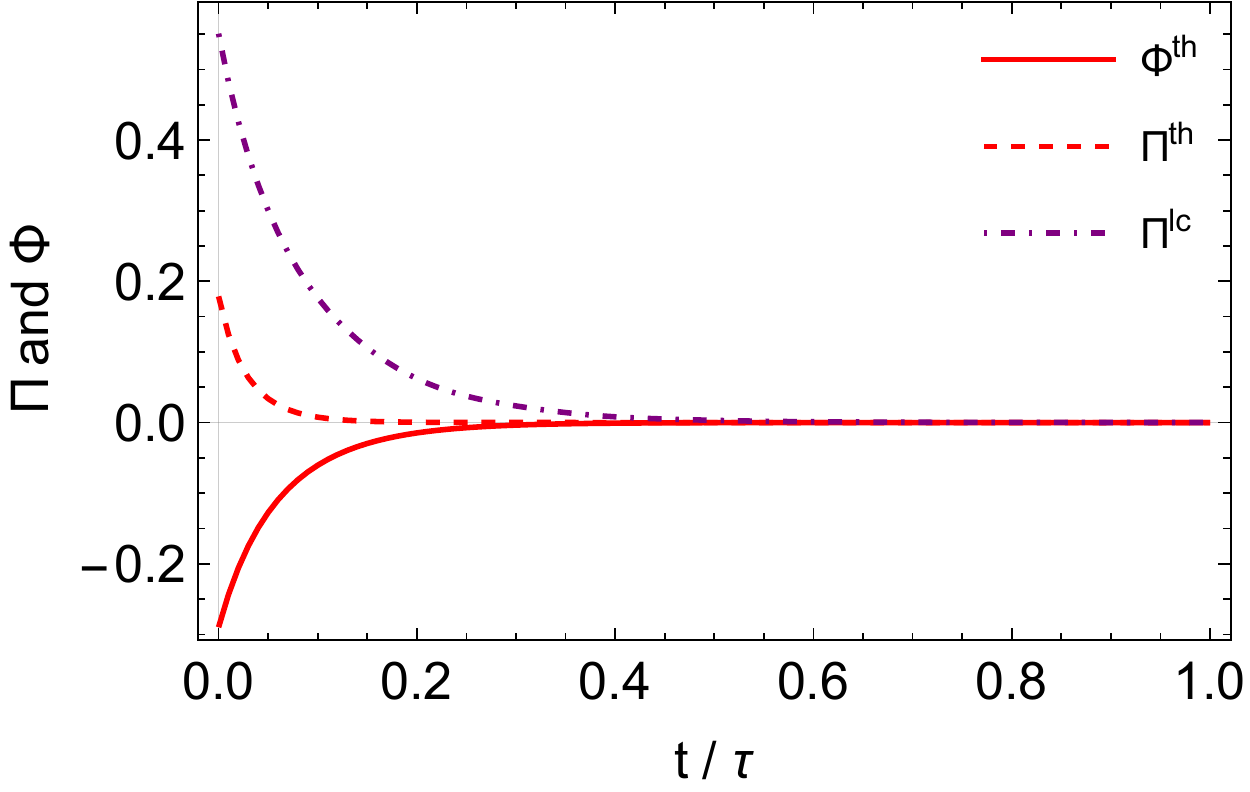}
    \end{minipage}
    \hfill%
    \begin{minipage}[t]{0.32\textwidth}
    \centering
        \centering
        \includegraphics[width=\textwidth]{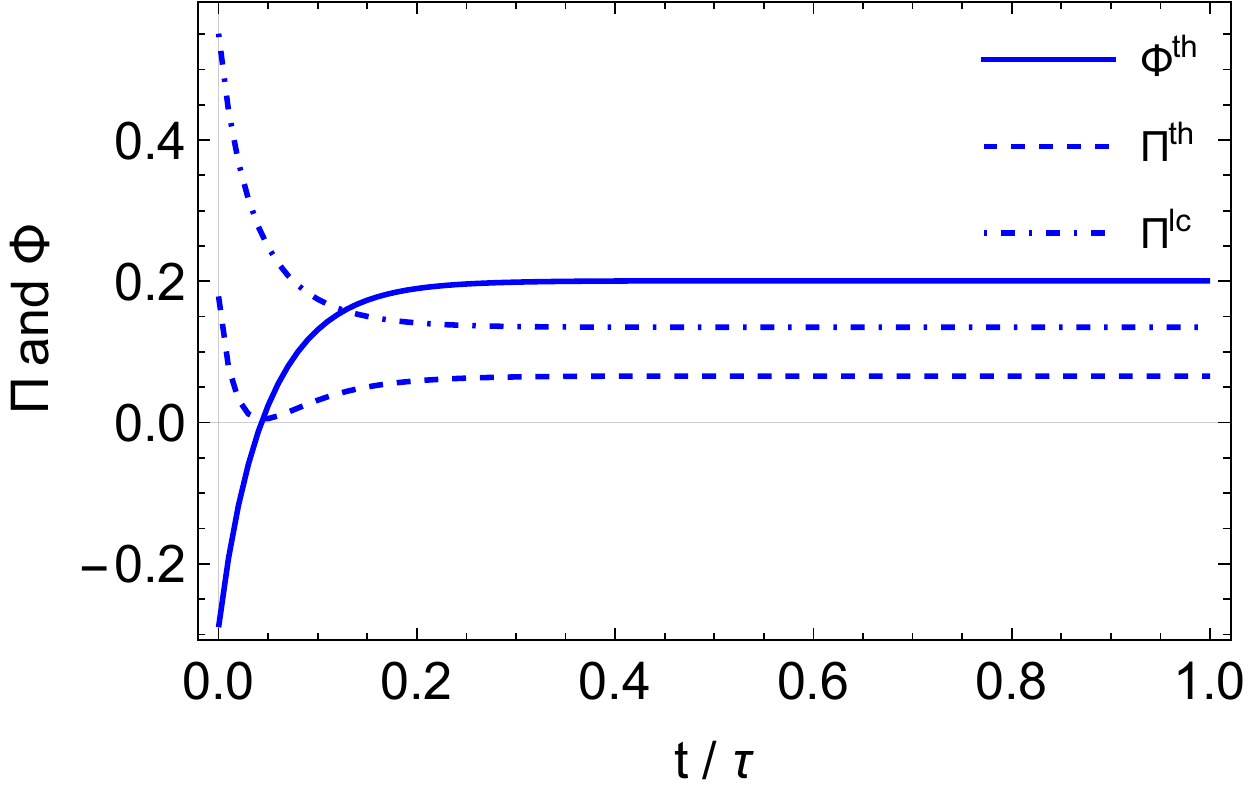}
    \end{minipage}
\caption{Study of a $4$-level spin system ($j=3/2$) subjected to different driving processes. Panels {\bf (a)}, {\bf (b)} and {\bf (c)} show respectively the evolution of the populations of the system under the action of only the thermalization, the localisation, and both dissipators together. The corresponding Wehrl entropies for each process are shown in Panel~{\bf (d)}. The  entropy production rates for processes with single dissipators are shown in Panel~{\bf (e)}, while those for the case with the combined action of both the dissipators are shown in Panel~{\bf (f)}.}
	\label{fig:thermodirectmethod}
\end{figure*}    

To understand the influences of the two dissipators, we consider the study case of a $4$-level spin system, whose Hamiltonian $H_\text{s}$ is given by
\begin{equation}\label{equ:systemhamiltonian1}
H_\text{s} = \omega J_z,
\end{equation}
where $J_z$ is the spin operator along $z$ direction.
The initial state is prepared to be the thermal state $\rho_i=\frac{e^{-\beta H_\text{s}}}{Z(\omega,\beta)}$ with $Z(\omega,\beta)=\sum_i e^{-\beta e_i}$ being the partition function, and $\{e_i\}$ are the eigenvalues of the system Hamiltonian. The system is initially prepared with inverse temperature $\beta=2/\hbar\omega$. In the following, we consider the processes of thermalisation and localisation separately, then we study their combined action.

\subsubsection{Thermalisation}
The system is attached to a thermal environment at the inverse temperature $\beta_\text{B}= 1/\hbar\omega$. This leads to a heating process where the system is asymptotically led to the thermal state
$\rho_\text{th}=\frac{e^{-\beta_\text{B} H_\text{s}}}{Z(\omega,\beta_\text{B})}$ with respect to the environmental temperature. The thermal dissipator is given in \eq{equ:thermomastereqsc} and we set $\gamma/\omega = 0.5$. 
The corresponding evolution of the populations of the system is shown in \subf{fig:thermodirectmethod}{a}. The increase of temperature leads to a system with less distinctive population distribution. Such process also changes the Wehrl entropy $S_{\cal Q}^\text{th}$ of the system, which is calculated in \eq{equ:wehrlentropy} and its evolution -- always increasing -- is shown in \subf{fig:thermodirectmethod}{d}. The decomposition of the Wehrl entropy rate into the entropy production rate $\Pi$ and the entropy flux rate $\Phi$ is shown in \subf{fig:thermodirectmethod}{e}. They can be respectively calculated from Eq.~\ref{equ:thermoentropyfluxrate}, and \eq{equ:thermoentropyproductionrate}. The trend of $\Pi^\text{th}$ shows that the production of irreversible entropy of the system is slowing down, while that of $\Phi$ indicates that the flux of entropy from the environment to the system is decreasing as the system is getting thermalised. One can see that the irreversible entropy and the entropy flux pace differently during the growth of the Wehrl entropy. Indeed,  the irreversible entropy rate $\Pi^\text{th}$ is quickly suppressed while the entropy flux rate $\Phi^\text{th}$ lasts for a longer time. For longer times,  both the rates go to zero, indicating that the state is fully thermalised. 

\subsubsection{Localisation}
Given any initial state, the localisation dissipator in \eq{equ:localmastereqsc} destroys the coherence of the system in positions, namely in the $J_x$ basis. Since the system Hamiltonian in \eq{equ:systemhamiltonian1} does not commute with $J_x$, their combined action  drive the system into a mixed state. 
We start with the same initial state considered in the previous case, and set the localisation parameter to  $\Lambda / \omega = 0.5$.
The corresponding evolution of the populations of the system is shown in \subf{fig:thermodirectmethod}{b}, where the final state is found to be the completely mixed state $\rho_\text f=\iden/4$, which  denotes  that  any  coherence  in  the  system  is  destroyed. The evolution of the Wehrl entropy $S_{\cal Q}^\text{lc}$ for this process 
is shown in \subf{fig:thermodirectmethod}{d}, and it is increasing asymptotically towards the value that corresponds to the thermal state at infinite temperature.
For this process, we have that the entropy flux rate $\Phi^\text{lc}$ is null. Thus, the Wehrl entropy only rate consists in the irreversible entropy production rate $\Pi^\text{lc}$. Its evolution in time is shown in \subf{fig:thermodirectmethod}{e}. Moreover, we notice that, although we took $\gamma=\Lambda$, the irreversible entropy production rates of the localization and thermalization process are not equal but following relation holds $\Pi^\text{lc}>\Pi^\text{th}$. 

\begin{figure*}
    \centering
    \begin{minipage}[b]{0.32\textwidth}
        \textbf{(a)}
    \end{minipage}%
    \hfill%
    \begin{minipage}[b]{0.32\textwidth}
        \textbf{(b)}
    \end{minipage}
    \hfill%
    \begin{minipage}[b]{0.32\textwidth}
        \textbf{(c)}
    \end{minipage}\\[1ex]
    \begin{minipage}[t]{0.32\textwidth}
        \centering
        \includegraphics[width=\textwidth]{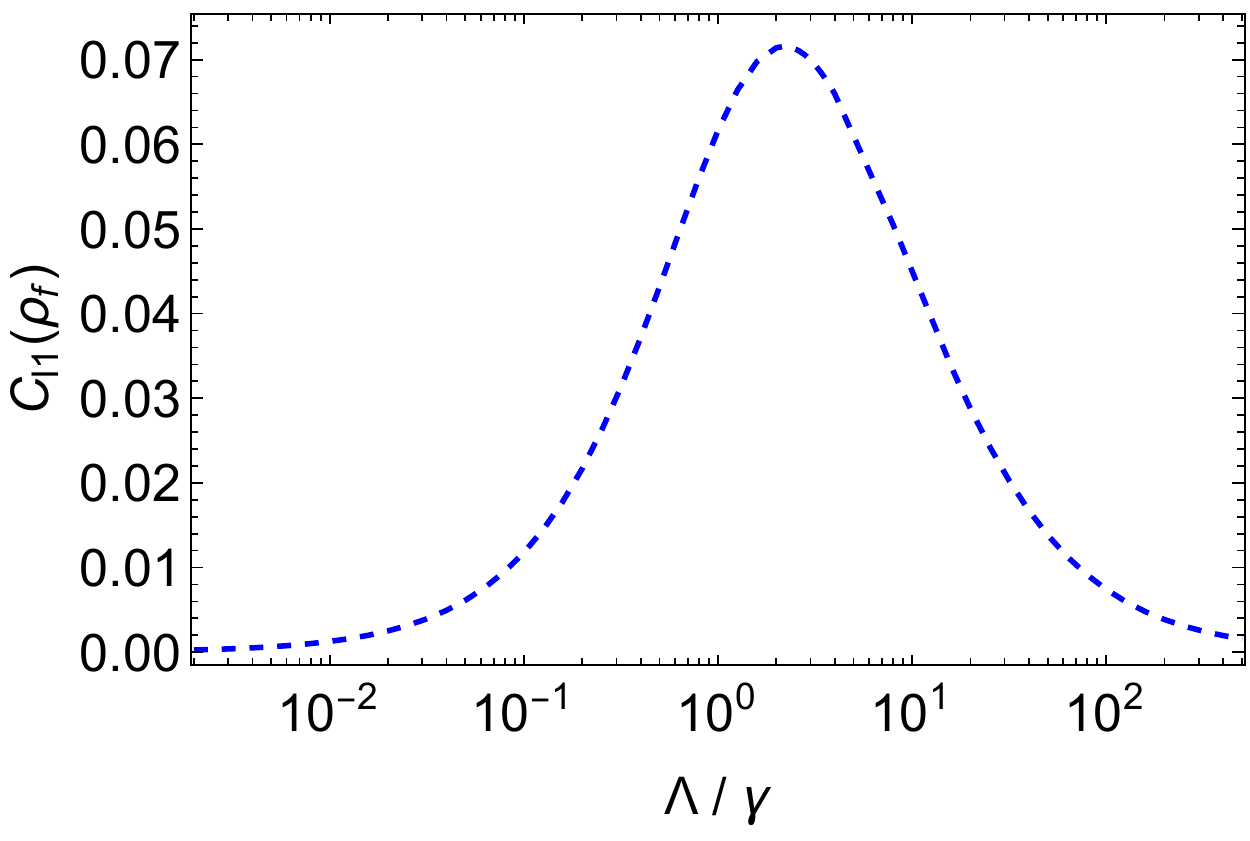}
    \end{minipage}%
    \hfill%
    \begin{minipage}[t]{0.33\textwidth}
    \centering
        \centering
        \includegraphics[width=\textwidth]{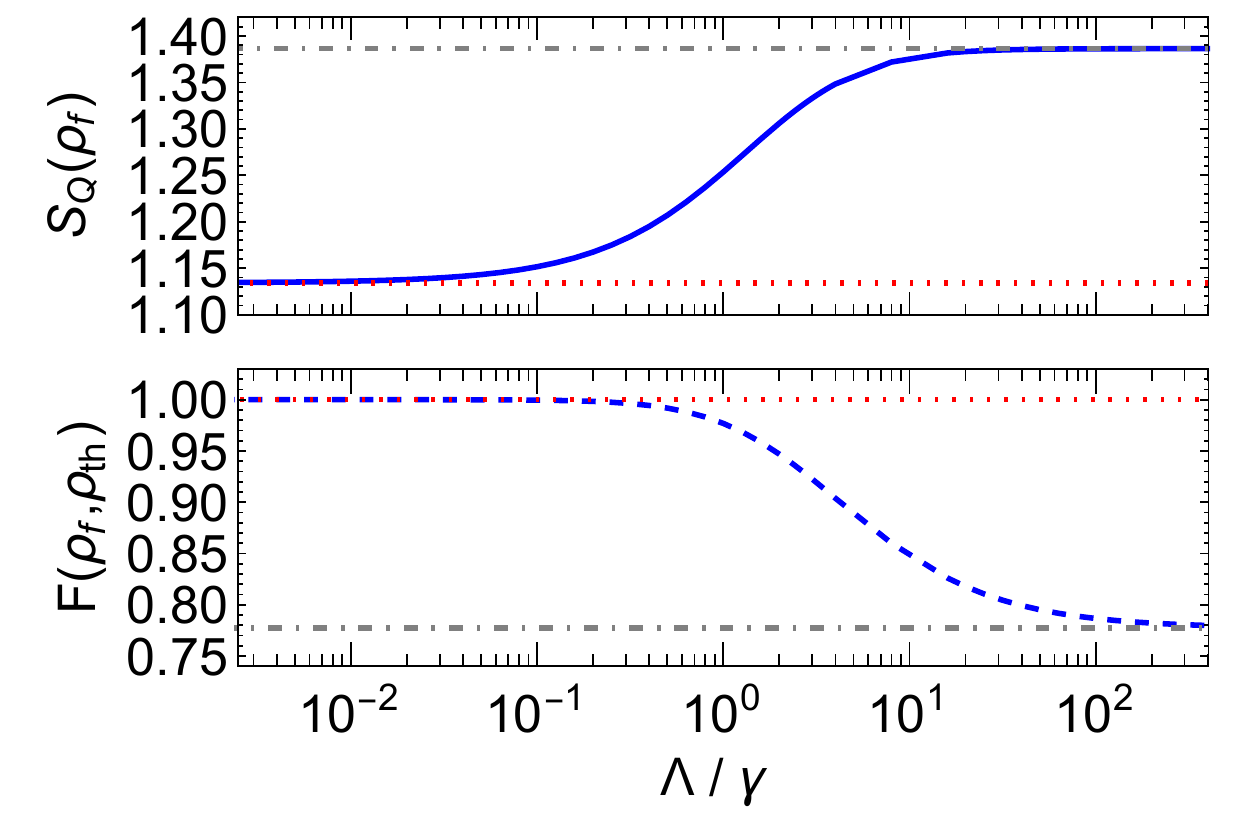}
    \end{minipage}
    \hfill%
    \begin{minipage}[t]{0.32\textwidth}
    \centering
        \centering
        \includegraphics[width=\textwidth]{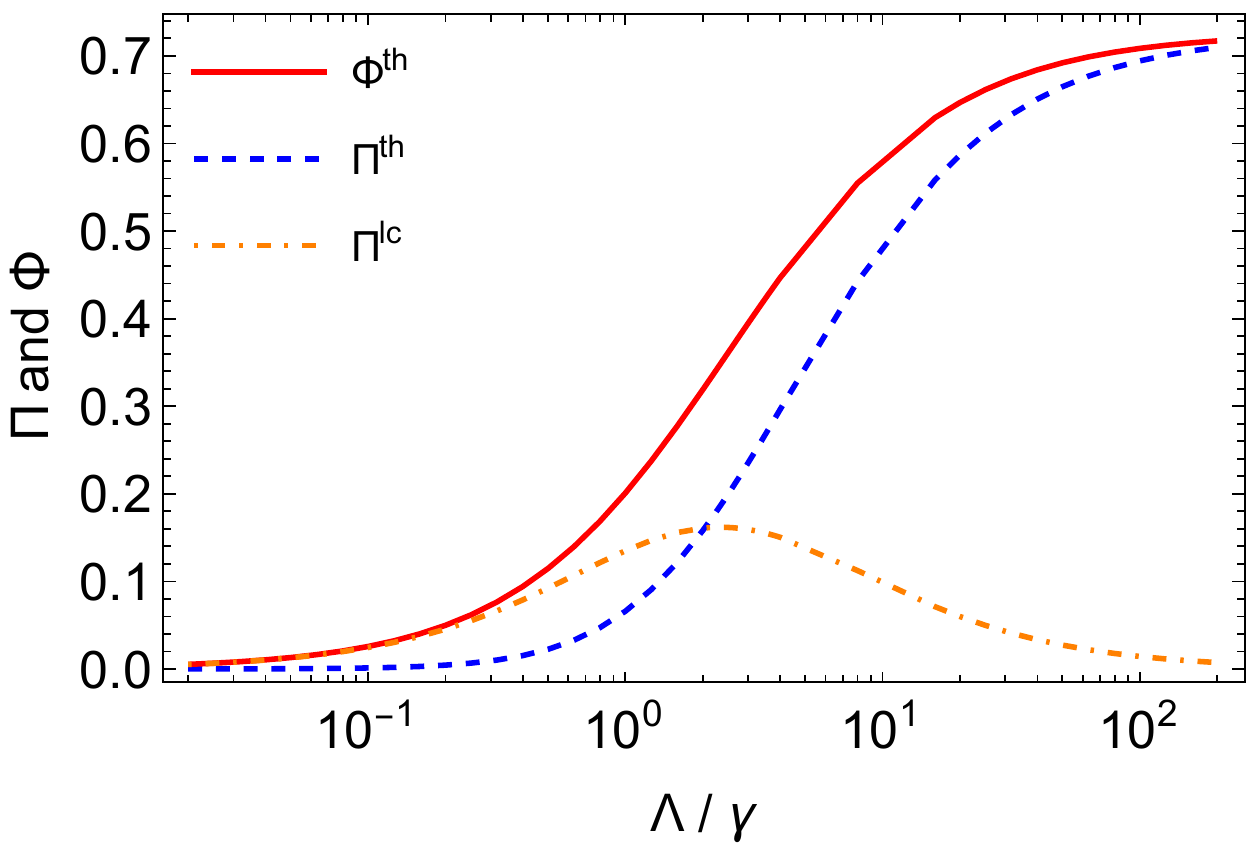}
    \end{minipage}
	\caption{Characteristics of the final non-equilibrium steady state given different coupling strengths $\gamma$ and $\Lambda$. Here we set $\gamma/\omega=0.5$ and we change the value of $\Lambda$.  Panel~{\bf {(a)}} shows the coherence in the final state, Panel~{\bf {(b)}} shows the Wehrl entropy $S_{\cal Q}$ of the system and quantifies the distance between the final state from the two-dissipator case and the final state from the thermalisation-only case in terms of the fidelity $F(\rho_\text{f},\rho_\text{th})$ with respect to the thermal state $\rho_\text{th}$. Here, the red dotted and the gray dot-dashed lines are references to the thermal state $\rho_\text{th}$ and localised state $\rho_\text{lc}$ respectively. Panel~{\bf {(c)}} shows the corresponding entropy production rates at the end of different processes, for which \eq{equ:ratesrelation} always holds true. }
	\label{fig:characteristicsofsteadstate}
\end{figure*}

\subsubsection{Combined dynamics}

Here we consider the dynamics of the system that is subject to the combined action of the thermalisation and the localisation dissipators. We consider the previous settings, namely the couplings are $\gamma/\omega=\Lambda/\omega=0.5$ and the inverse temperature for the environment is $\beta_\text{B}=1/\hbar \omega$.
The corresponding evolution of the populations of the system is shown in \subf{fig:thermodirectmethod}{c}. Since two dissipators  drive asymptotically the system to  different  states, this will eventually end in a non-equilibrium steady state. This is manifested by the evolution of its Wehrl entropy and production rates. One can see in  \subf{fig:thermodirectmethod}{d} that the Wehrl entropy grows quickly at the beginning of the process, due to collective action of both dissipators, and then it stabilises to an average value between those of the two single processes. The entropy production rates of the combined dynamics are shown in \subf{fig:thermodirectmethod}{f}. In contrast to the previous two cases with only one dissipator acting, here the rates no longer approaching zero but a stable positive value, underling that the system has reached a non-equilibrium steady state. Moreover, since the localisation dissipator is driving the system toward an infinite temperature state, the system is overheated with respect to the temperature of the environment. Thus, the environment plays the role of cooling the system, which is indicated by the transition of the entropy flux rate $\Phi^\text{th}$ from  negative (when is still heating the system) to positive values. On the other hand, the irreversible entropy rates $\Pi^\text{th}$ and $\Pi^\text{lc}$ remain positive throughout of the process. Once a stable situation is attained, the following relation holds
\begin{equation}\label{equ:ratesrelation}
    \Pi^\text{th}+\Pi^\text{lc}-\Phi^\text{th}=0,
\end{equation}
which indicates that the system has stabilised at a non-equilibrium steady state. Here, the system is overheated, and feeds to the environment the excessive entropy produced by the positive irreversible entropy productions of the two dissipators.



We now deepen the study of the non-equilibrium steady state under the action of the combined dynamics by changing the relative strength of the coupling constants of the two dissipators.  Without loss of generality, we set again $\gamma / \omega=0.5$ and let the value of $\Lambda$ change.  The first quantity of interest is the 
coherence, which is quantified with a $l_1$ measure in the basis of $J_z$ as \cite{Baumgratz_2104} 
\begin{equation}\label{equ:l1coherence}
    C_{l_1}(\rho) = \sum_{j\neq k}|\rho_{jk}|.
\end{equation}
We observe that at the end of the process there is still present some coherence in the final non-equilibrium steady state as it is shown in \subf{fig:characteristicsofsteadstate}{a}. Here, the two dissipators drive the system in different bases: the system is prepared in basis of $J_z$ and the $D_\text{lc}[\rho]$ localizes the system in basis of $J_x$; due to their non-commutativity, the system is driven out of equilibrium during the process, thus maintaining some coherence. Such a coherence would disappear if only the localisation dissipator $D_\text{lc}[\rho]$ acts on the system. However, when both dissipators are present, the thermalisation dissipator $D_\text{th}[\rho]$ prohibits the system from being fully localised in basis $J_x$, yet fails to fully thermalise the system, resulting in a final non-equilibrium steady state containing non-zero coherence. We see the coherence peaks at $\Lambda/\gamma \approx 2$, and the coherence disappears when $\Lambda=0$ or $\infty$, where the dynamics is dominated by one of the two dissipators.

We proceed by quantifying the influences of the thermalisation and localisation dissipators in driving the system towards the fully thermalised state $\rho_\text{th}$ and the fully localised state $\rho_\text{lc}$, respectively. To quantify this driving, we employ fidelity, which is defined as
\begin{equation}
    F(\rho,\sigma) = \left( \tr{\sqrt{\sqrt{\rho}\sigma\sqrt{\rho}}} \right)^2.
\end{equation}
In particular, 
\subf{fig:characteristicsofsteadstate}{b} shows the fidelity $F(\rho_\text{f},\rho_\text{th})$ between the final and the thermal state, with referencing lines that represent $\rho_\text{th}$ (the red dot line) and $\rho_\text{lc}$ (the gray dot-dashed line). On the other hand, we also demonstrate the Wehrl entropy of the final state w.r.t different coupling strengths, compared with Wehrl entropies of $\rho_\text{th}$ and $\rho_\text{lc}$. From both plots, one can observe a non-linear transition as $\Lambda$ increases. 

Finally, in \subf{fig:characteristicsofsteadstate}{c} we compare the entropy production rates for different values of the ratio of the couplings. We analyse them at the end of the 
simulations, when the system reaches the non-equilibrium final state and the relation in \eq{equ:ratesrelation} holds true.
As the ratio $\Lambda/\gamma$ increases, one can observe an increase of the irreversible entropy production rate $\Pi^\text{th}$ and the flux rate $\Phi^\text{th}$ from the thermalisation dissipator. This is due to the localization dissipator, which pushes the system toward an ``infinite temperature'' state. On the other hand, one has a peak of the irreversible entropy production rate $\Pi^\text{lc}$ from the localisation dissipator at $\Lambda/\gamma\approx 2$, for whose value the final state displays the maximum coherence, as it is shown in  \subf{fig:characteristicsofsteadstate}{a}.

\subsection{Quantum oscillator with non-Gaussian potential}

\begin{figure*}[tb!]
	\centering
    \begin{minipage}[b]{0.32\textwidth}
        \textbf{(a)}
    \end{minipage}
    \hfill
    \begin{minipage}[b]{0.32\textwidth}
        \textbf{(b)}
    \end{minipage}
    \hfill
    \begin{minipage}[b]{0.32\textwidth}
        \textbf{(c)}
    \end{minipage}\\[1ex]
     \begin{minipage}[t]{0.32\textwidth}
        \includegraphics[width=\textwidth]{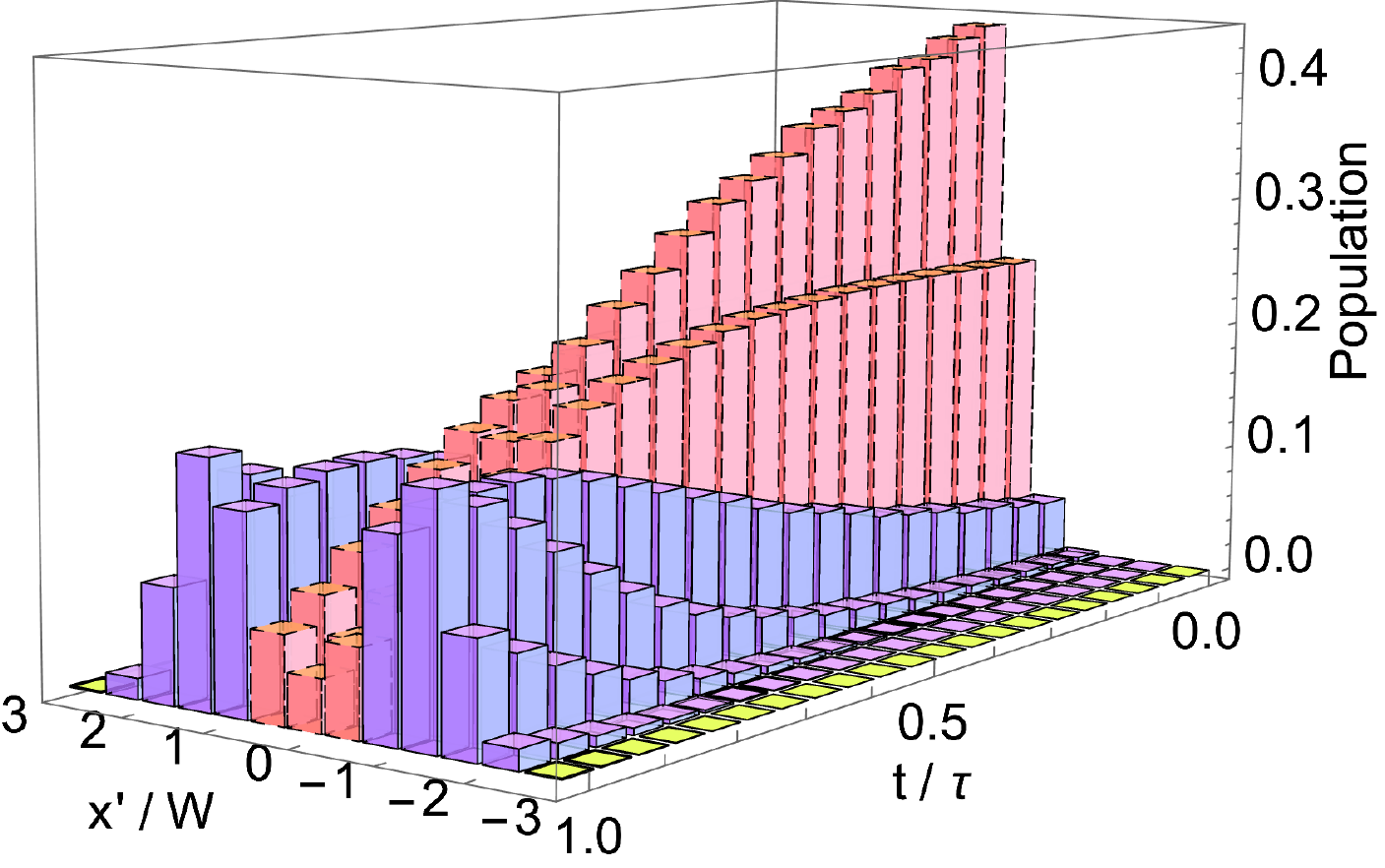}
    \end{minipage}
    \hfill
    \begin{minipage}[t]{0.32\textwidth}
        \includegraphics[width=\textwidth]{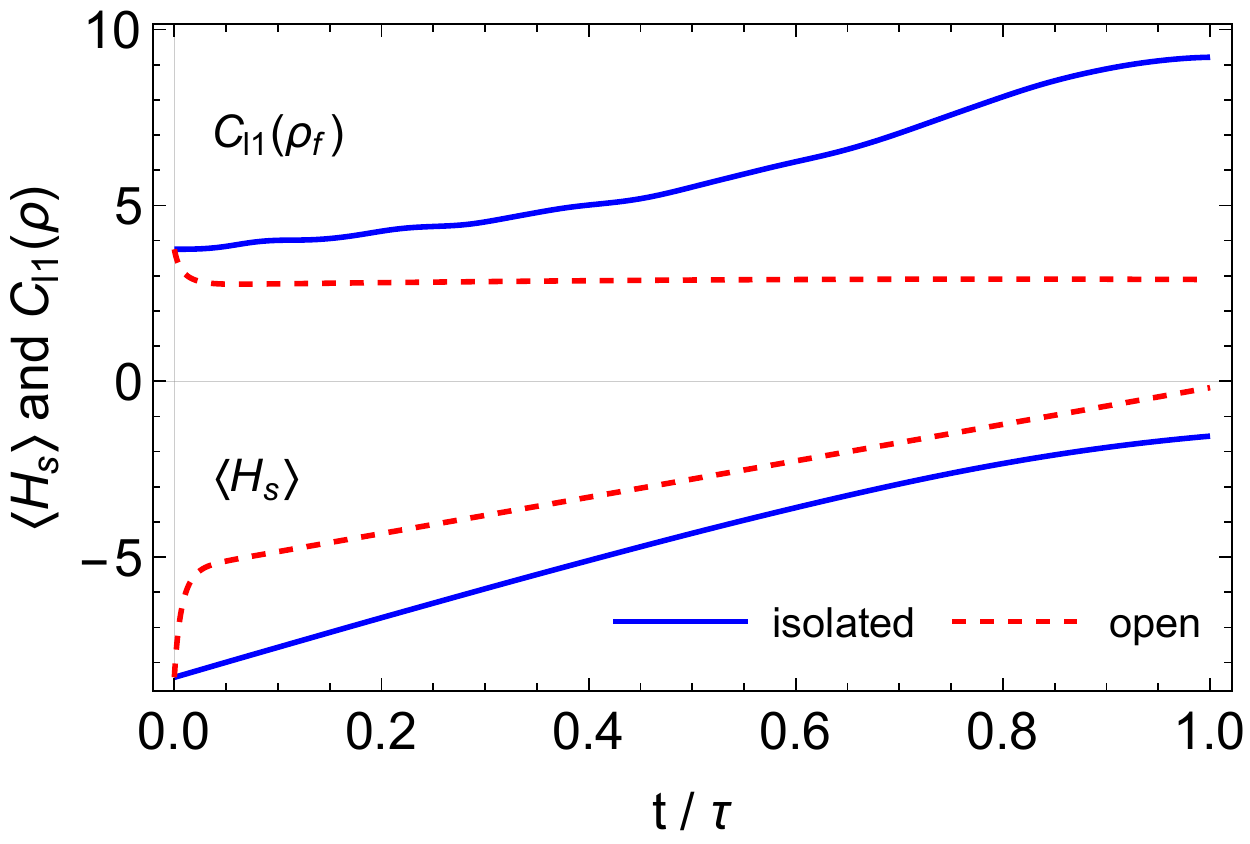}
    \end{minipage}
    \hfill
    \begin{minipage}[t]{0.32\textwidth}
        \includegraphics[width=\textwidth]{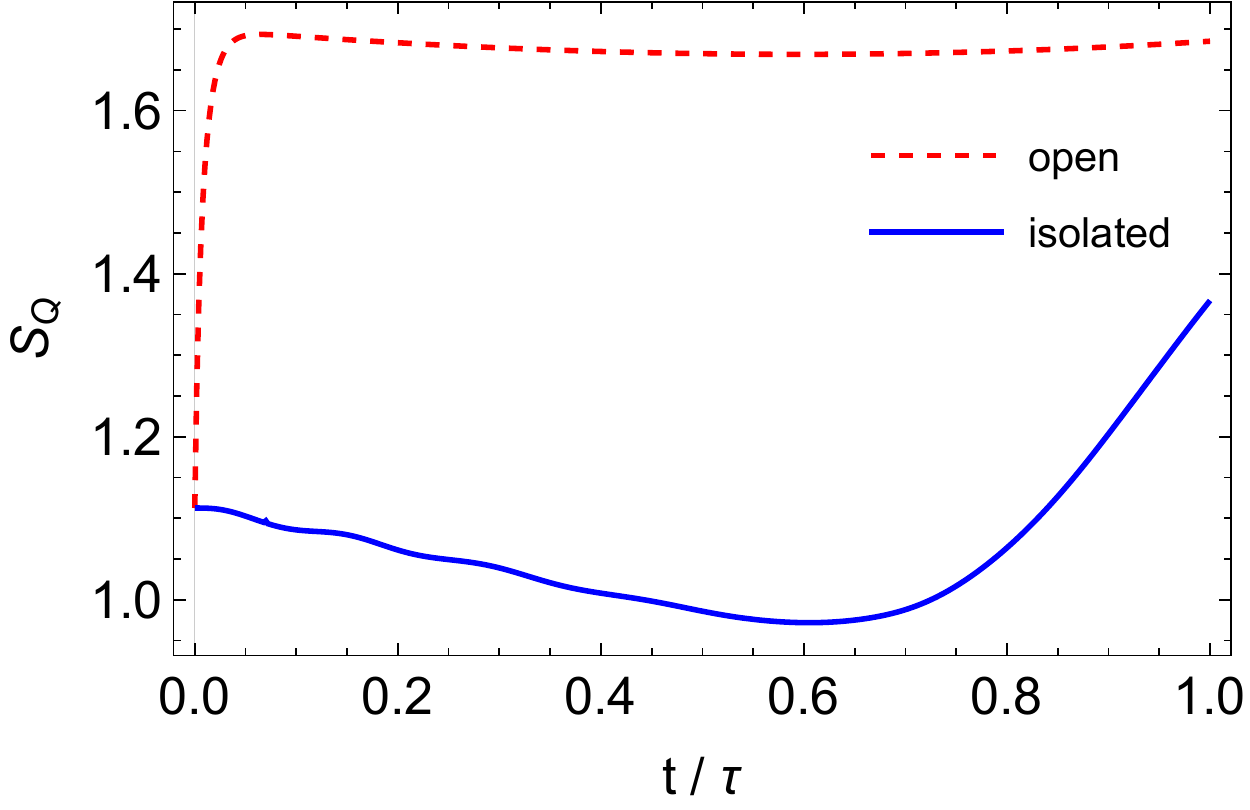}
    \end{minipage}\\[2ex]
 \begin{minipage}[b]{0.32\textwidth}
        \textbf{(d)}
    \end{minipage}%
    \hfill%
    \begin{minipage}[b]{0.32\textwidth}
        \textbf{(e)}
    \end{minipage}
    \hfill%
    \begin{minipage}[b]{0.32\textwidth}
        \textbf{(f)}
    \end{minipage}\\[1ex]
     \begin{minipage}[t]{0.32\textwidth}
        \centering
       \includegraphics[width=\textwidth]{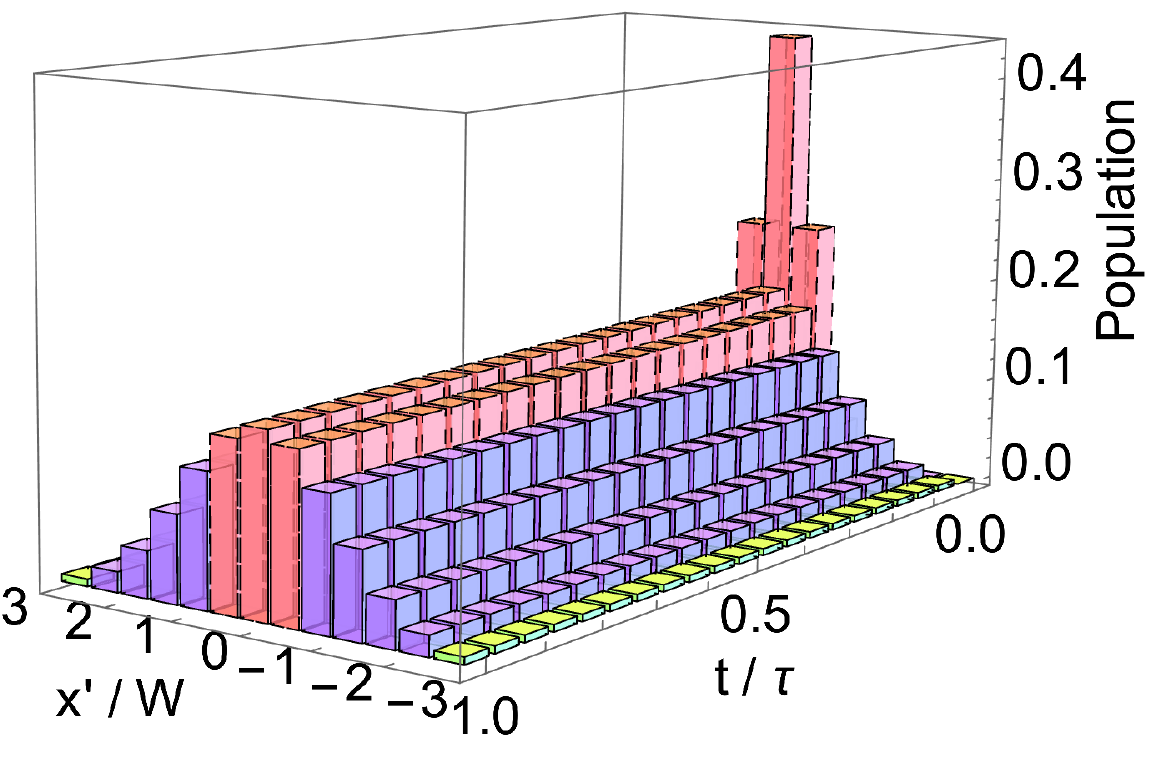}
    \end{minipage}%
    \hfill%
    \begin{minipage}[t]{0.32\textwidth}
        \centering
        \includegraphics[width=\textwidth]{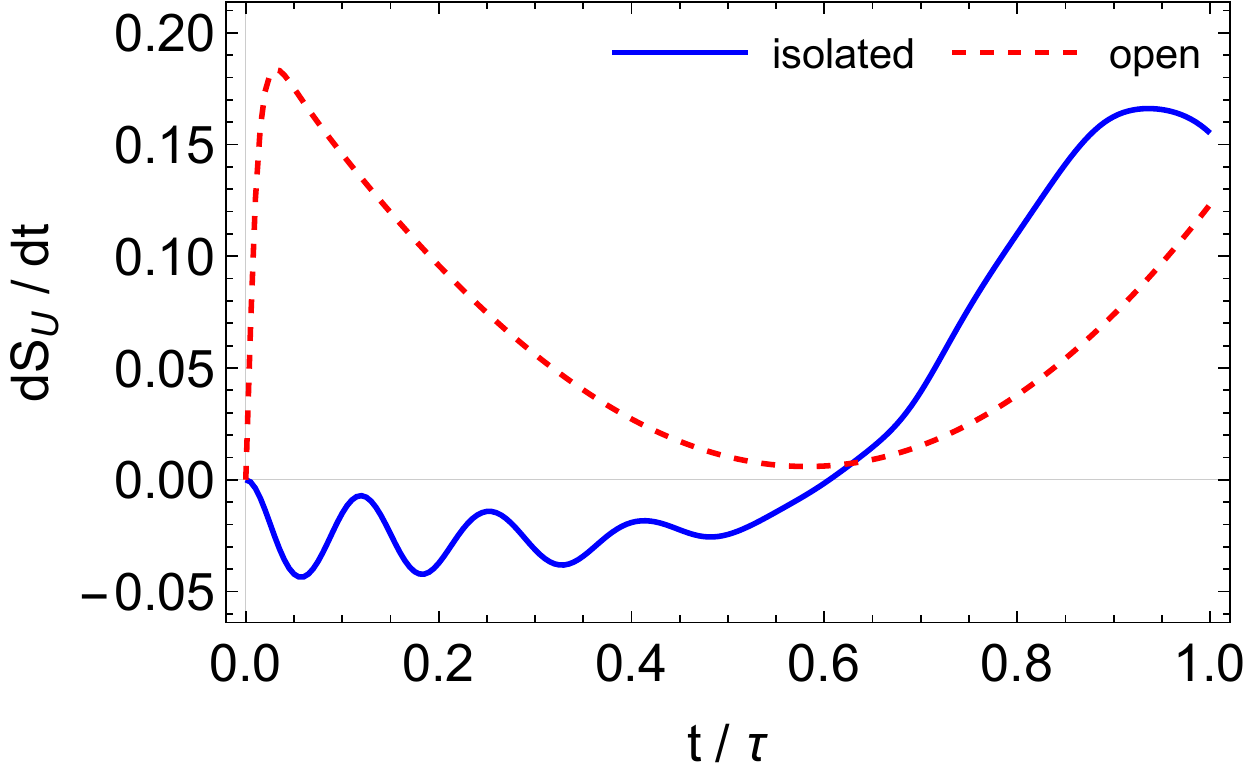}
    \end{minipage}
    \hfill%
    \begin{minipage}[t]{0.32\textwidth}
        \centering 
        \includegraphics[width=\textwidth]{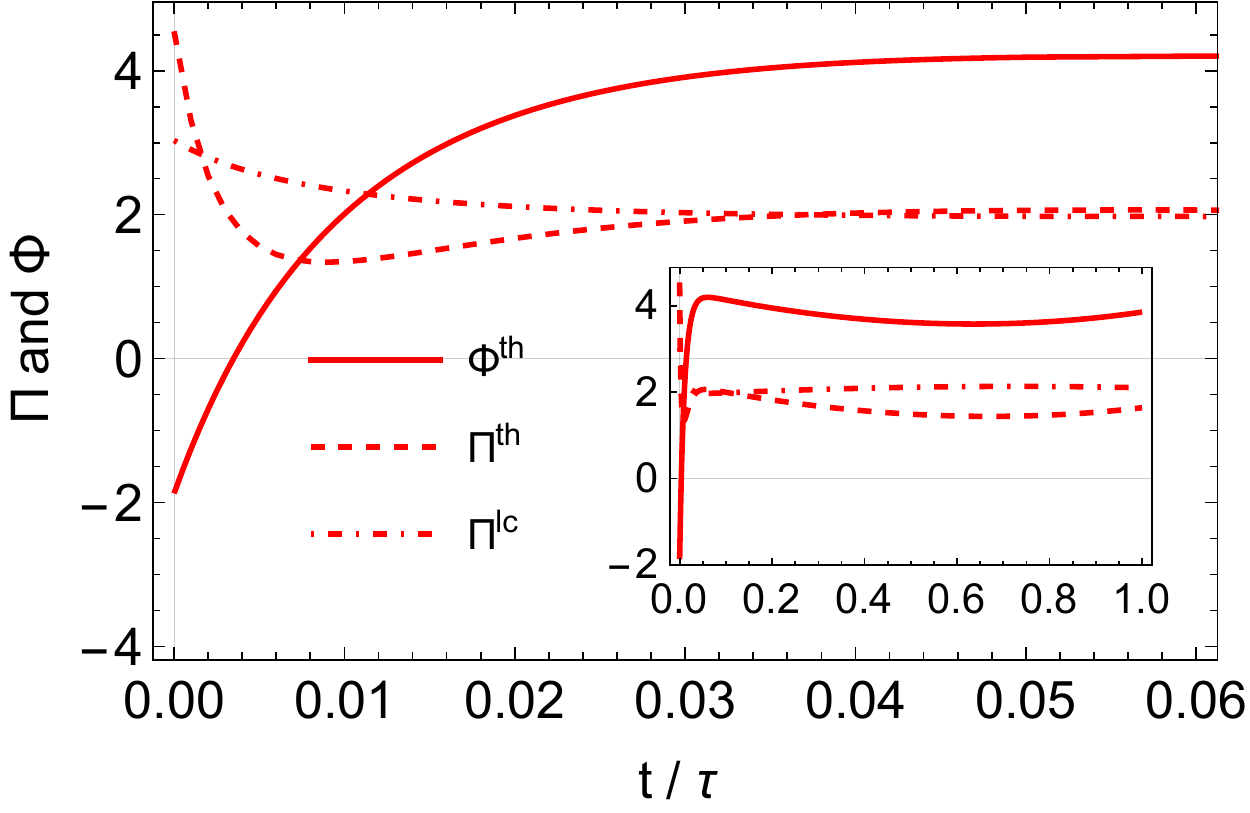}
    \end{minipage}
	\caption{Dynamics of a 25-level spin system with a Gaussian-to-double-well potential. Panels~{\bf (a)} and {\bf (d)} show the population changes for the isolated and the combined dynamics in the spin basis. The respective features are shown in the other panels, where we used the blue and the red colors to identify respectively the isolated case and that of the combined dynamics. Panel~{\bf (b)} shows the mean energy and the coherence, while the Wehrl entropy is displayed in Panel~{\bf (c)}. Panel~{\bf (e)} shows the unitary contribution to Wehrl entropy production rate, while Panel~{\bf (f)} exhibits the corresponding irreversible entropy production rates $\Pi^\text{th}$, $\Pi^\text{lc}$ and the entropy flux rate $\Phi^\text{th}$ for the combined dynamics at the beginning of the dynamics. Their entire evolution down to time $t=\tau$ is illustrated in its inset.  }
	\label{fig:closeddoublewell}
\end{figure*} 

We now focus on the quantum oscillator with the time-varying Hamiltonian redefined in Eqs.~\eqref{equ:qohamiltonianinqsc} and \eqref{equ:doublepotentiainqsc}, with dimension $N=25$ and Taylor expansion up to the order $\kappa=15$. The initial state is prepared as the thermal state with respect to $H_\text{s}^\text{sp}(0)$ 
at $\beta = 2/\hbar \omega$ (we remand below for the analysis with different initial states). Here we analyze two different situations:  the case of the isolated system, which is not as trivial as in the case of the spin system, and the open dynamics under the influence of both the dissipators. 

\subsubsection{Isolated dynamics}
In the case of the isolated dynamics, we are interested in verifying the effective splitting of the wavefunction distribution when going from the Gaussian-harmonic potential to the double-well. Thus, we first verify, in the $J_{x'}$ basis, the proper separation of the initial population in the two peaks as the dynamics takes place. 
The plot in \subf{fig:closeddoublewell}{a} shows the effects of the time-varying potential in the population distribution. Since at $t=0$ the system is prepared at a low temperature, the population initially resides mostly in the lowest energy level, which is localized in a region being small with respect to the distance of the double well [cf.~\subf{fig:doublewellpotentialplot}{a}]. Then, the population divides in two peaks corresponding to the two minima of the double-well potential at $t=\tau$.
Since the double-well potential has a shallower depth and larger eigenenergies [cf.~\subf{fig:doublewellpotentialplot}{c}], the energy of the system increases during the transition, as described by \subf{fig:closeddoublewell}{b}. Also the coherence of the system, which is quantified by \eq{equ:l1coherence} in basis of $J_{x'}$, grows in time. Indeed, the double-well potential, by splitting the population in two localized regions, generates coherence between these two that results in its increasing behaviour reproduced in \subf{fig:closeddoublewell}{b}.
Finally, we show the Wehrl entropy of the system and its rate respectively in \subf{fig:closeddoublewell}{c} and \subf{fig:closeddoublewell}{e}. They both capture the non-equilibrium features caused by the change in the potential.
{Initially, the small deformation of the potential from its quadratic behaviour pushes the state sightly out of the equilibrium. Consequently, the  Wehrl entropy rate presents small oscillations, which are captured in the  first half of the evolution presented in \subf{fig:closeddoublewell}{e}. In the second half of such evolution, the potential starts to present the central peak characterising the double-well. Subsequently, the  non-equilibrium driving becomes stronger and leads to a stronger increases of the Wehrl entropy of the system and its rate.}

\subsubsection{Combined dynamics}
We now consider the combined dynamics where the system is under the action of both the thermalisation and the localisation dissipators. Here, we set the coupling rates to $\gamma/\omega=\Lambda/\omega=0.5$ and the inverse temperature for the environment $\beta_\text{B}=1/\hbar \omega$, which are the same values considered in the study case considered in \sect{subsect.spin}.
The main effect of dissipators is to wash away all the quantum features of the system, and thus it tries to nullify the action of the transition to the double-well. In \subf{fig:closeddoublewell}{d}, we show the evolution of the population under this combined dynamics. As one can see, the population is modified only in the first moments of the evolution, where the distribution slightly broadens without splitting in the expected two peaks. The first part of the evolution corresponds to the initial increase of the energy driving the system in a higher temperature state by the two dissipators (cf.~\subf{fig:closeddoublewell}{b}). After this initial change, the distribution in position appears as ``frozen'' for the rest of the dynamics.
{It is shown in \subf{fig:closeddoublewell}{b}, the energy of system increases with an initial surge driven by 
the two dissipators until the corresponding inverse temperature $\beta_\text{B}$ is reached. Then, the energy continues to increase but at a lower rate only due to the change of the potential. Indeed,  the combined action of the two dissipators does not contribute to the energy change: although the localisation heats the system, the thermalisation cools it back to $\beta_\text{B}$.} Alongside with this, the coherence becomes smaller, which reflects the illegible effect of the potential on the population distribution, until it reaches a constant value as a result of the action of the two dissipators driving the system in two different basis (the same was noticed in the spin study case, cf. \subf{fig:characteristicsofsteadstate}{a}).
The Wehrl entropy of the system  and its rates
are shown  respectively in \subf{fig:closeddoublewell}{c} and \subf{fig:closeddoublewell}{f}. Similarly as the spin situation, the Wehrl entropy increases due to the thermalisation and the localisation. The latter overheats the system with respect to the thermal dissipator, so that the entropy flux rate $\Phi^\text{th}$ changes from being negative to positive. Eventually, the system reaches a non-equilibrium steady state for which the decomposed rates then follow \eq{equ:ratesrelation} since 
the unitary rate $\D S_\mU/\D t$ is negligible when compared to the other rates.

\begin{figure}[htb]
    \centering
    \begin{minipage}[b]{\linewidth}
        \textbf{(a)}
    \end{minipage}\\[1ex]
    \begin{minipage}[t]{0.88\linewidth}
        \centering
        \includegraphics[width=\linewidth]{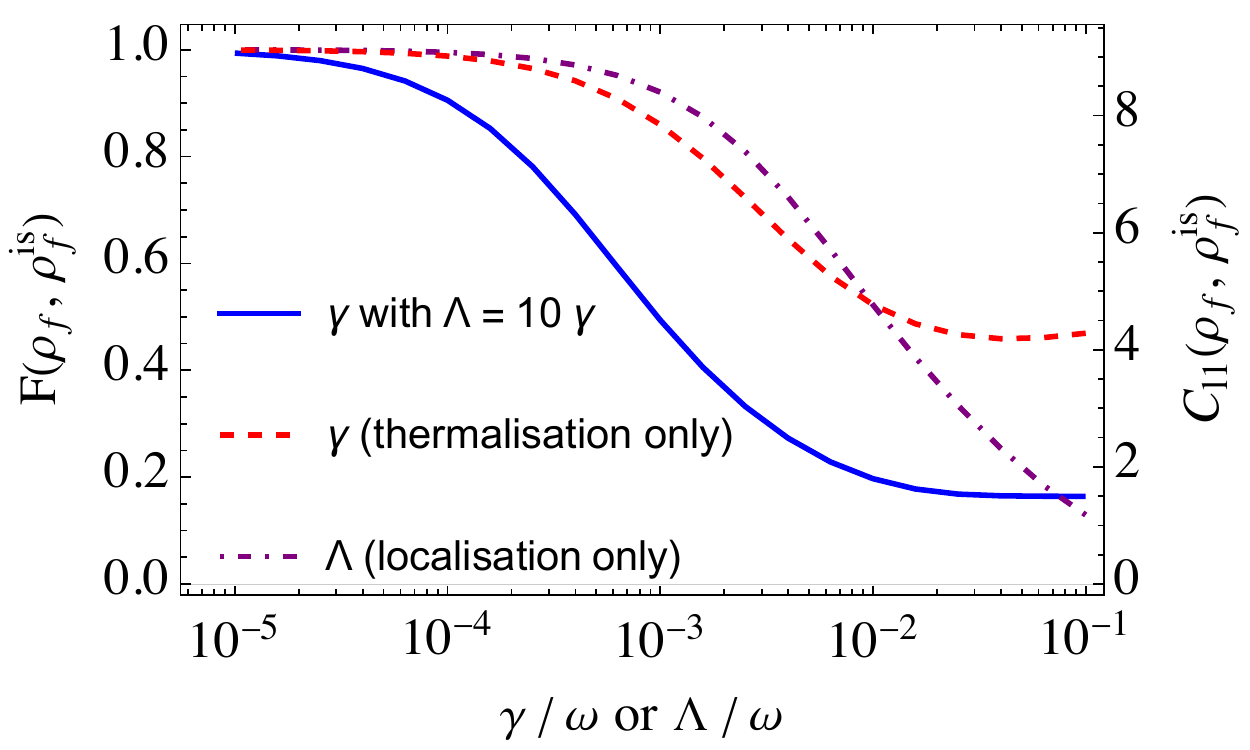}
    \end{minipage}\\[2ex]
    \begin{minipage}[b]{\linewidth}
        \textbf{(b)}
    \end{minipage}\\[1ex]
    \begin{minipage}[t]{0.78\linewidth}
    \centering
        \centering
        \includegraphics[width=\linewidth]{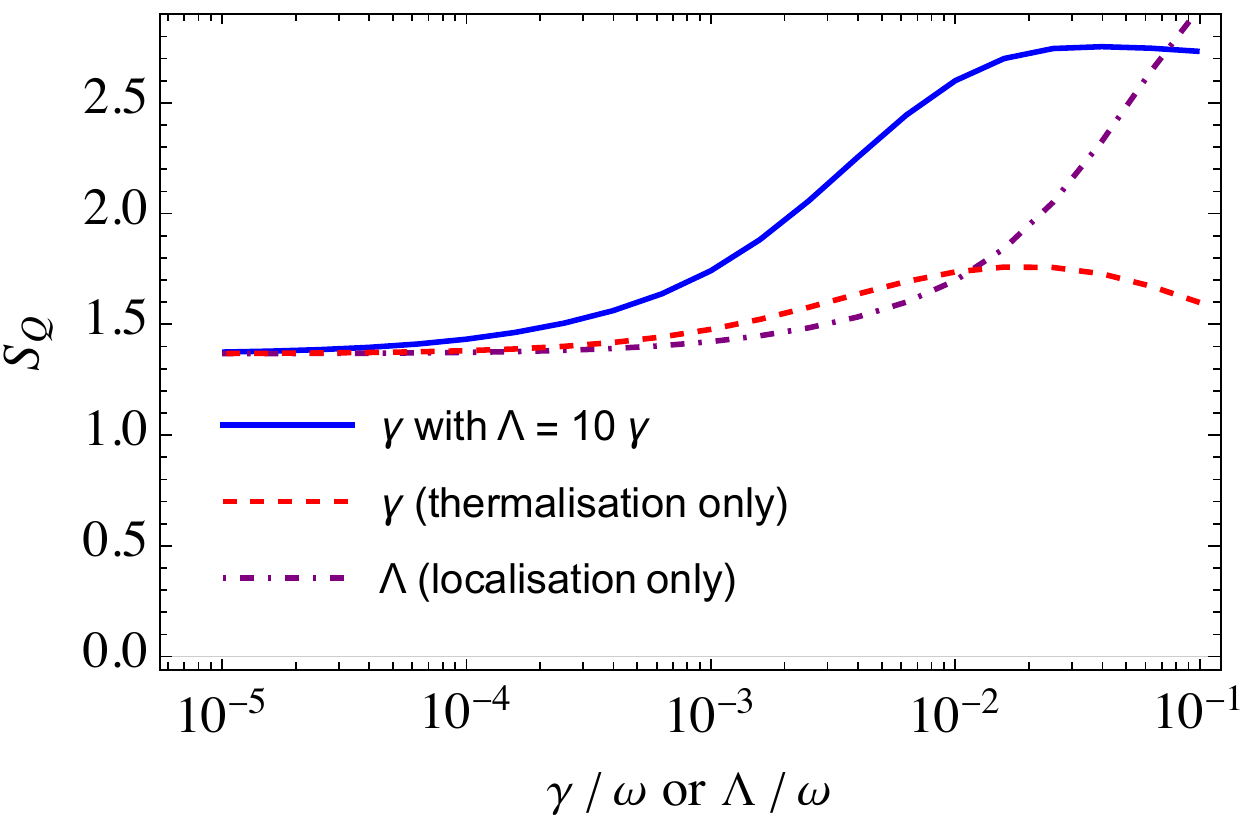}
    \end{minipage}
    \hspace{0.3cm}
	\caption{
	Analysis of the final state for different values of the coupling constant for three scenarios: the case with only thermalization is represented with a dashed red line, that with only localization with a purple dot-dashed line, while that with both the dissipators with the fixed ratio $\Lambda/\gamma=10$ with the continuous blue line.
	Panel~{\bf (a)} shows the fidelity between the final states of different coupling strengths and the final state from the isolated system, which is the target state. Panel {\bf (c)} displays the Wehrl entropy of the final states. }
	\label{fig:fidvsgammalambda}
\end{figure}

\paragraph{Different coupling regimes} 
The action of the two dissipators has prevented the generation of a population in position characterized by two peaks relative to the minima of the double-well potential. We thus proceed to analyse which can be the 
suitable coupling strengths for which such a two-peak distribution can still be observed even though the presence of the two dissipators.
Here, we consider three scenarios: the one where one has only thermalisation, that with only  localisation, and that with both the dissipators being present. 
Today, optical levitation experiments performed in low vacuum allow to reach a condition where photon recoil is significantly stronger then damping. We thus explore this regime fixing $\Lambda/\gamma=10$ for the rest of these simulations. See \sect{sec:experimentperspective} for detailed discussions. 

We consider two figures of merit to make a full comparison of these scenarios. The first one is the fidelity $F(\rho_\text{f},\rho_\text{f}^\text{is})$ between the non-equilibrium steady state for the considered dynamics and the final state for the isolated dynamics, and it is shown in \subf{fig:fidvsgammalambda}{a}. One can infer that the previous simulation (where $\Lambda/\omega=\gamma/\omega=0.5$) reproduced a low-fidelity final state. To improve such a situation, one is required to reduce the values of $\Lambda$ and $\gamma$. In  particular, for being able to reach a fidelity of $F>80\%$, one needs to set the coupling strength $\gamma$ or $\Lambda$ to approximately $10^{-3}\omega$ in the case of a scenario with a single dissipator. The requirement becomes instead more stringent, with $\gamma/\omega\approx10^{-4}$ and $\Lambda/\omega\approx10^{-3}$, when both the dissipators are involved in the dynamics. These values are close to the current state-of-art experimental setting (See \sect{sec:experimentperspective}). {Notably, the coherence $C_{l_1}$, which is defined in Eq.~\eqref{equ:l1coherence}, provides approximately the same behaviour as the fidelity, up to a proportionality constant. We display its scaling in the right vertical axis of \subf{fig:fidvsgammalambda}{a}.}
Finally, we consider the Wehrl entropy of the final state as the second figure of merit, which is shown in \subf{fig:fidvsgammalambda}{c}. When considering the two single-dissipator cases, one notices that the thermalisation has a milder action on the entropy compared to that of the localisation.
Indeed, the thermalisation dissipator heats the system up to the inverse temperature $\beta_\text{B}$, but not beyond that. On the other side, the localisation dissipator, which can be treated as a bath at infinite temperature, keeps heating the system. Consequently, also its action in terms of entropy is stronger. In the case where both the dissipators are considered with the fixed relation of $\Lambda=10\gamma$, one can identify two different regimes: up to $\gamma/\omega\sim10^{-2}$ both dissipators slowly heat the system; beyond that value, the thermalisation works as a cooling mechanism, in contrast to the localisation that keeps heating the system. This results in a increasing behaviour of the Wehrl entropy up to $\gamma/\omega\sim10^{-2}$, after which it is slowly descending.


\begin{figure*}[htb]
	\centering
    \begin{minipage}[b]{0.32\textwidth}
        \textbf{(a)}
    \end{minipage}
    \hfill
    \begin{minipage}[b]{0.32\textwidth}
        \textbf{(b)}
    \end{minipage}
    \hfill
    \begin{minipage}[b]{0.32\textwidth}
        \textbf{(c)}
    \end{minipage}\\[1ex]
     \begin{minipage}[t]{0.32\textwidth}
        \includegraphics[width=\textwidth]{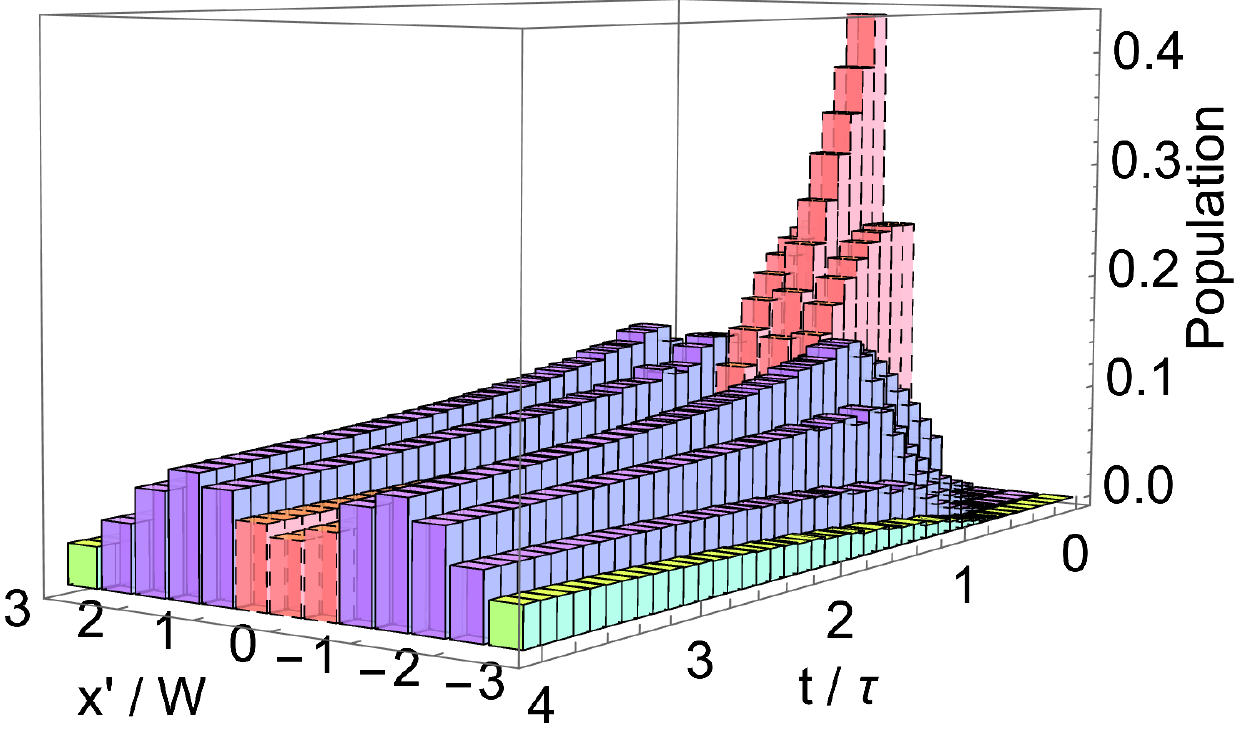}
    \end{minipage}
    \hfill
    \begin{minipage}[t]{0.32\textwidth}
        \includegraphics[width=\textwidth]{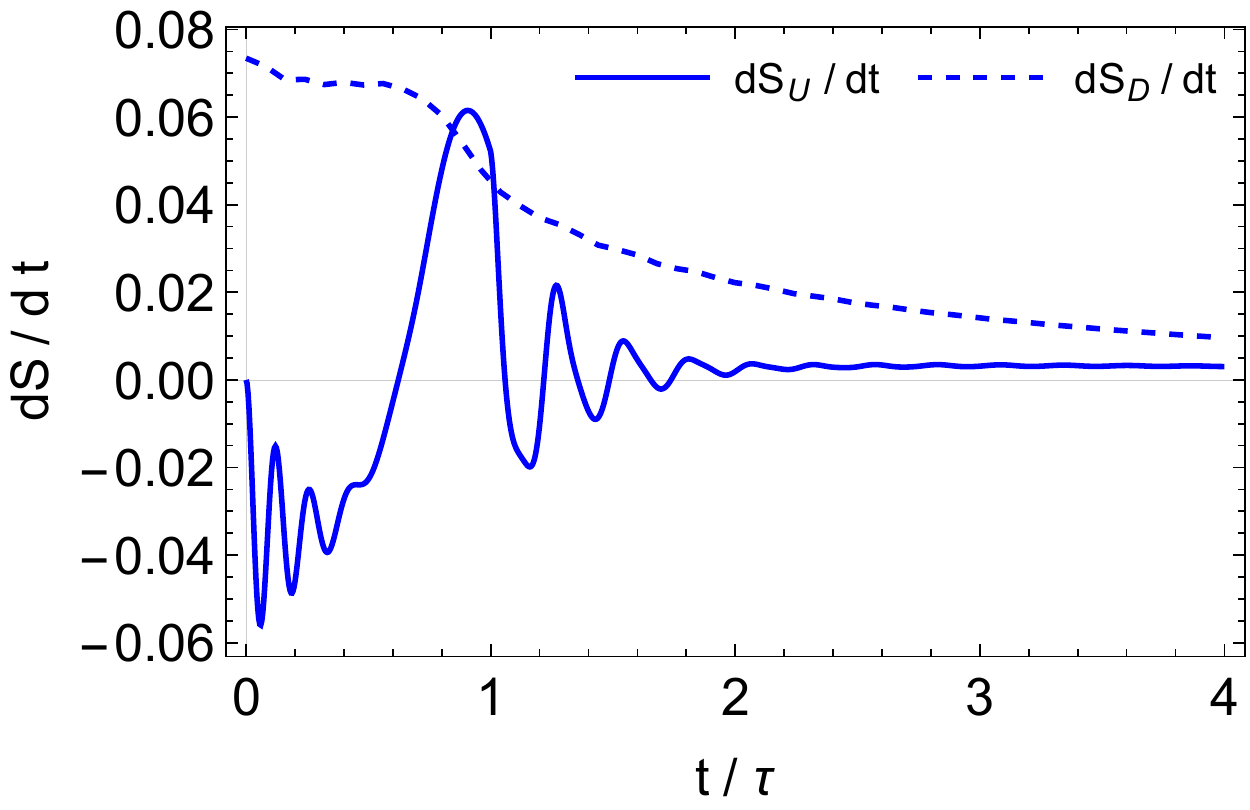}
    \end{minipage}
    \hfill
    \begin{minipage}[t]{0.32\textwidth}
        \includegraphics[width=\textwidth]{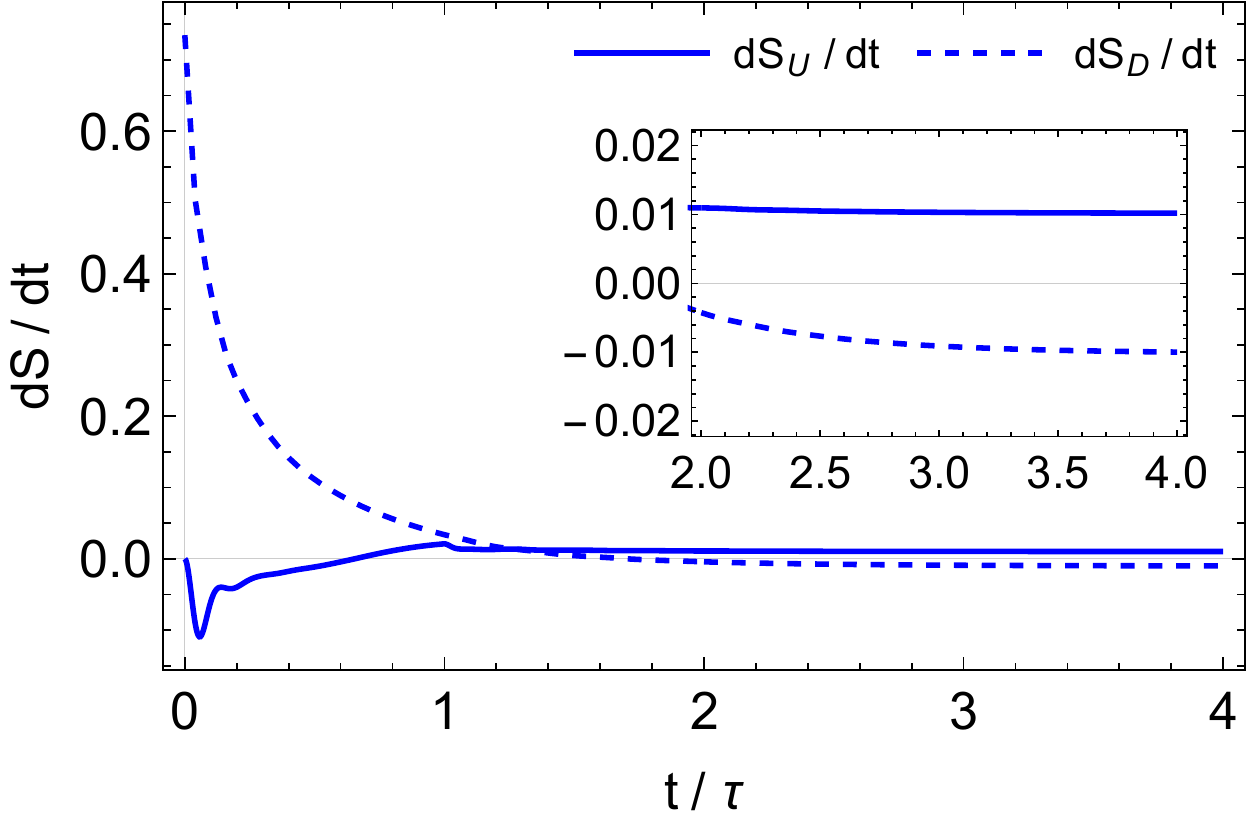}
    \end{minipage}\\[2ex]
 \begin{minipage}[b]{0.32\textwidth}
        \textbf{(d)}
    \end{minipage}%
    \hfill%
    \begin{minipage}[b]{0.32\textwidth}
        \textbf{(e)}
    \end{minipage}
    \hfill%
    \begin{minipage}[b]{0.32\textwidth}
        \textbf{(f)}
    \end{minipage}\\[1ex]
     \begin{minipage}[t]{0.31\textwidth}
        \centering
       \includegraphics[width=\textwidth]{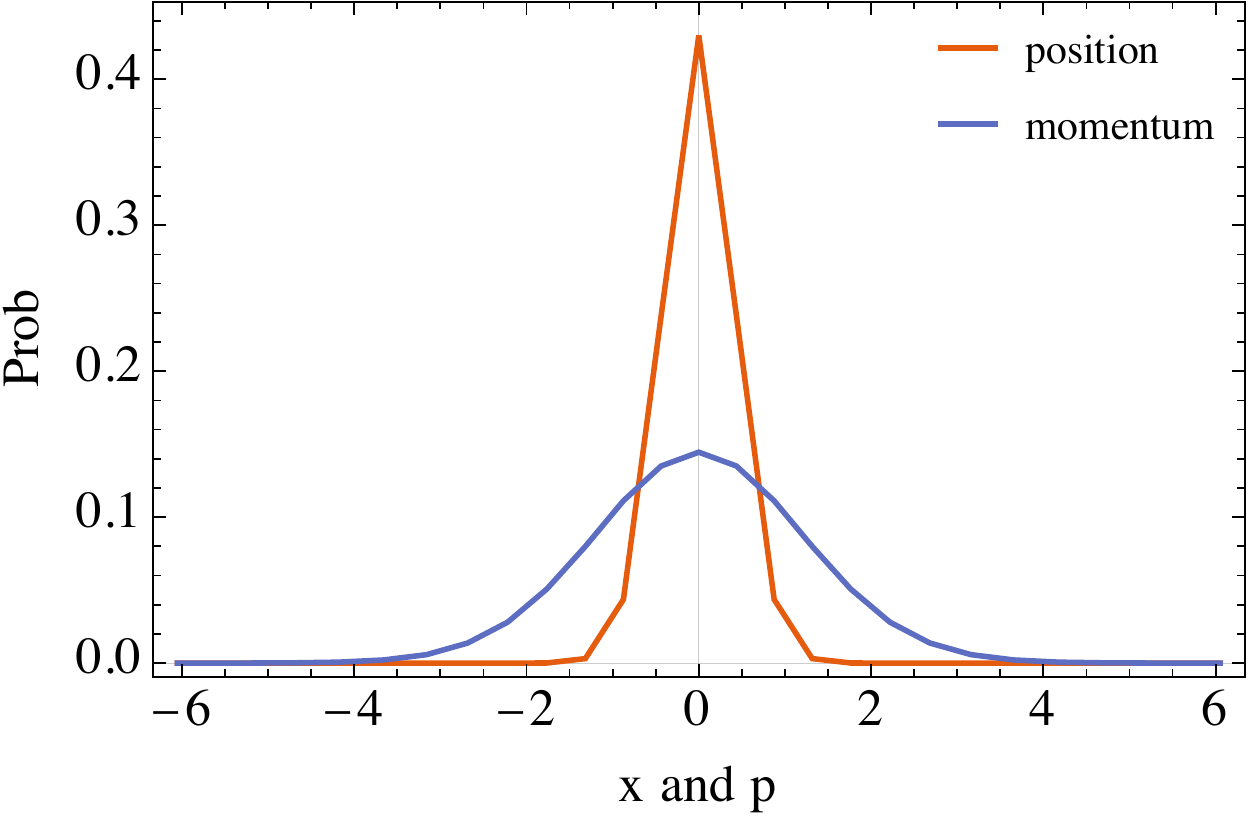}
    \end{minipage}%
    \hfill%
    \begin{minipage}[t]{0.33\textwidth}
        \centering
        \includegraphics[width=\textwidth]{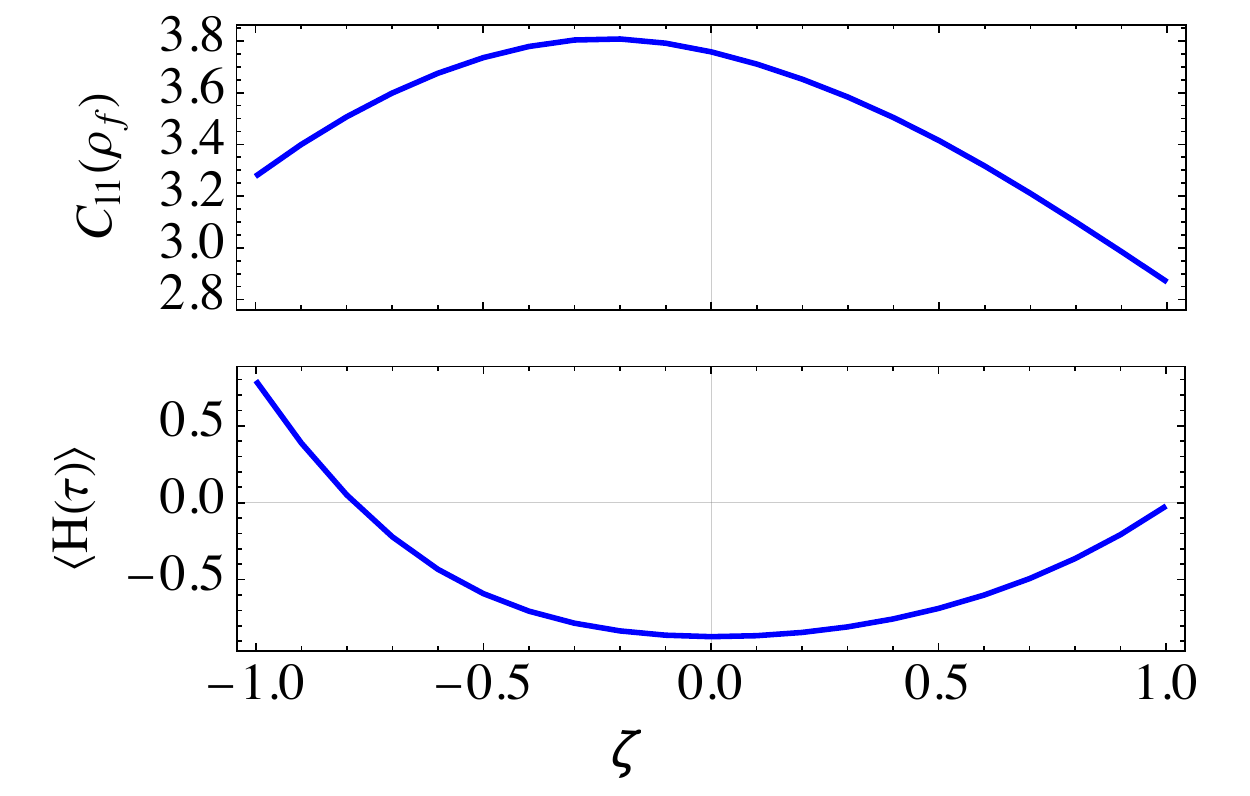}
    \end{minipage}
    \hfill%
    \begin{minipage}[t]{0.32\textwidth}
        \centering 
        \includegraphics[width=\textwidth]{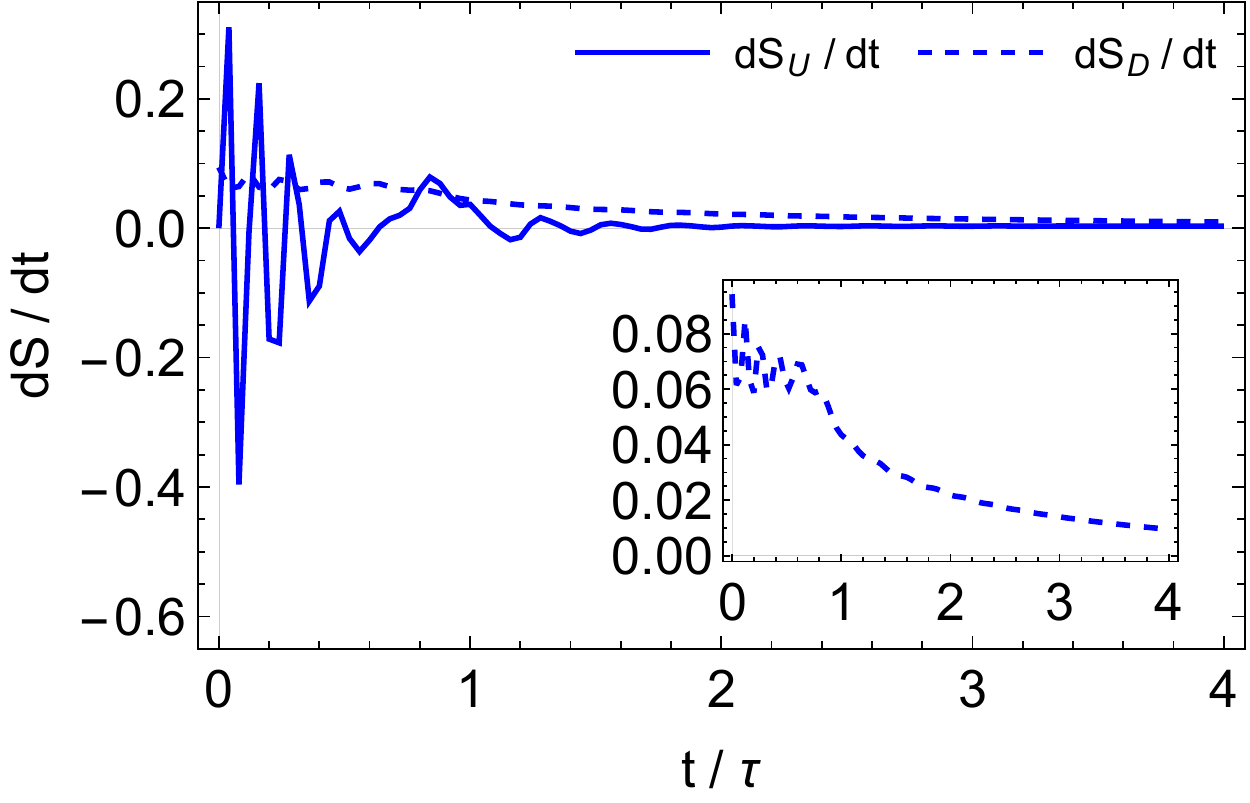}
    \end{minipage}
	\caption{Evolution with smaller coupling rates and different squeezed initial states. Here, we set $\Lambda=10\gamma$. Panel~{\bf (a)} shows the evolution of the distribution in positions with coupling strengths $\gamma/\omega=10^{-3}$ and evolution time $t/\tau\in[0,4]$. Panels~{\bf (b)} and {\bf (c)}, respectively for $\gamma=10^{-3}$ and  $\gamma=10^{-2}$, compare the unitary contribution $\frac{\D S_\mU}{\D t}$ (solid line) and the dissipative contribution $\frac{\D S_\mD}{\D t}$ (dashed line) to the Wehrl entropy rate $\frac{\D S_Q}{\D t}$. Panel~{\bf (d)} shows the position and momentum distribution of the initial state, and Panel~{\bf (e)} shows the final system energy and coherence given the preparation of the initial state at different squeezing values. For the case where the maximum coherence is achieved when $\zeta=-0.2$. Panel~{\bf (f)} shows the dynamics of the corresponding Wehrl entropy rates.}
	\label{fig:furtherstudy}
\end{figure*}

\paragraph{Low coupling regime.}
\label{sec:furtherstudy}
Given the results of the analysis above, we consider a low coupling regime that can potentially showcase stronger quantum features at the end of the protocol, and maintain them also after it. In particular, we set $\Lambda = 10 \gamma$ and consider two values for $\gamma$: case 1. $\gamma/\omega = 10^{-3}$, and case 2. $\gamma/\omega = 10^{-2}$. For both cases, we extend the dynamics after the transformation protocol to $t=4\tau$. This gives an insight on the environmental effects on the system also after the change of the potential.

In the low-coupling regimes, the first thing to notice is that we observe again the double-well effect in the population. See for example the population of case 1. in \subf{fig:furtherstudy}{a} where one can see the double-peak distribution also for $t=4\tau$. In both cases, the thermodynamics of the system gets comparable influences both from the unitary dynamics due to the potential and the action of the environment. Indeed, the magnitude of
the unitary contribution $\frac{\D S_\mU}{\D t}$ and the dissipative one $\frac{\D S_\mD}{\D t}$ to the Wehrl entropy rate are compared in \subf{fig:furtherstudy}{b}  and \subf{fig:furtherstudy}{c} respectively for the case 1.~and 2.
 From these figures, one observes that, for larger the coupling strength, the difference between the unitary and the dissipative contributions gets larger. This is the reason why, in the previous simulation with $\gamma/\omega=0.5$, it was possible to neglect the unitary part and only investigate the dissipative contribution. Moreover, after the end of the transition to the double-well potential, the system is still in contact with the environment and it tends to the steady state with a rate depending on the values of the coupling strengths. Such a process suggests that the total rates of Wehrl entropy production is going to zero asymptotically. However, with smaller coupling strengths, one can no longer ignore the unitary contribution $\frac{\D S_\mU}{\D t}$. This causes an interesting situation, which is shown in \subf{fig:furtherstudy}{c}. The system tends to  a non-equilibrium steady state that is neither in equilibrium with the system Hamiltonian nor the environment. Indeed, one gets $\frac{\D S_\mU}{\D t}>0$ and $\frac{\D S_\mD}{\D t}<0$, which means that the system keeps producing entropy due to non-equilibrium with the the Hamiltonian, while in meantime it dissipates the entropy in the environment.

\paragraph{Different initial states}
Due to the Gaussian term in the Hamiltonian in Eq.~\eqref{equ:doublepotential}, the initial state of the single-well potential is naturally  slightly squeezed in position, as it is shown in \subf{fig:furtherstudy}{d}. Here, we want to quantify which are the effects of such a squeezing on the performance of the protocol under study. To this end, we consider other  initial states by varying the amount of initial squeezing and study if this can provide a better outcome. The squeezing operator is given by \cite{gerry_knight_2004} 
\begin{equation}
    S(\zeta) = \exp{\left(\frac{1}{2}[\zeta^\ast b^2 - \zeta b^{\dagger 2}]\right)}
\end{equation}
and the squeezed initial state is prepare by $\rho^\text{sq}_\text{in}=S(\zeta)\rho_\text{in}S(\zeta)^\dagger$ with $\zeta\in[-1,1]$. In particular, we consider a similar simulation as above with the values of the parameters fixed at $\gamma=10^{-3}$, $ \Lambda=10\gamma$.
{Firstly, we show in \subf{fig:furtherstudy}{e} the energy and coherence of the system at $t=\tau$ at different values of initial squeezing $\zeta$. Regardless of the quadrature being squeezed, the initial squeezing adds energy to the system. On the other hand, a suitable squeezing can increase the coherence. The maximum in coherence is given by $\zeta=-0.2$, which corresponds to a small squeezing in momentum. Taking such a squeezed state as the initial state, we simulate the system and compute the dynamics of unitary and dissipative Wehrl entropy rates, which are shown in \subf{fig:furtherstudy}{f}. Comparing them qualitatively to the non-squeezed case (cf.~\subf{fig:furtherstudy}{b}), the squeezing induces huge oscillations to the unitary Wehrl entropy rate, while it does not affect significantly the dissipative rate. After the deformation of the potential, both rates approach to zero as in the non-squeezed case.}

\section{Experimental perspective} \label{sec:experimentperspective}

The experimental implementation of time-controlled double-wells in the quantum regime is challenging. Yet, recent experiments with optically levitated nanoparticles provide the necessary ingredients. Quantum limited position readout and the preparation of nearly pure quantum states of motion \cite{Delic_2020, Magrini_2021, Tebbenjohanns_2021} as well as the dynamical shaping of double-well potentials \cite{Ciampini_2021} have been demonstrated.
Thus, the protocol discussed here is in principle implementable, once these two methods are combined. This requires  the implementation of existing state detection methods \cite{Magrini_2021, Tebbenjohanns_2021}  to the detection of the transversal motion \cite{Tebbenjohanns_2019}, where the double-well potential landscape is realized. 
Yet, the downside in optical levitation is the unavoidable decoherence due to recoil of optical tweezer photons, resulting predominantly in a localization term.  Additionally, thermalization with the environment may be induced by collisions with the surrounding gas molecules. 
In the following, we base the discussion on the currently most favorable parameters in this respect, achieved by the recent cryogenic implementation of an optical harmonic trap by Tebbenjohanns \textit{et al.}~\cite{Tebbenjohanns_2021}.

\paragraph{System and potential.} We consider a silica nanoparticle of radius $R=50$\,nm trapped in an dynamically controlled optical double well potential, created from a combination of a TEM$_{00}$ and TEM$_{01}$ optical mode \cite{Ciampini_2021}. The trapping laser has a wavelength of $\lambda=1550$\,nm. We assume a frequency of  $\omega=2\pi\times 54.9$\,kHz in Eq.~\eqref{equ:systemhamiltonian2}, which  corresponds to an effective optical trapping frequency of $2\pi\times 77.6$\,kHz when we add also the contribution from $H_\text{add}$ with $\alpha=1$. The timescale of the protocol $\tau$ is limited by the switching times of the light field, which can be several orders of magnitude faster than the mechanical frequency when done with electro-optic modulators. It can therefore be considered near instantaneous. Using a numerical aperture of the optical tweezer $N=0.75$ the beam waist is $W \approx 660$\,nm. Assuming equal restoring forces contributing to $H_\text{s}$ and $H_\text{add}$ we find  ${\cal E}=8 \times 10^5 \hbar \omega$ in Eq.~\eqref{equ:doublepotential}. 

\paragraph{Localisation dissipator.}

The major effect of localization is due to the scattering of tweezer photons from the levitated nanoparticle. We assume that this effect in the optical trap exceeds localization due to blackbody radiation, which is expected to be fulfilled in current experiments. The corresponding decoherence parameter at the maximum intensity of a Gaussian beam is given by \cite{Seberson_2020}:
\begin{equation}
    \Lambda=\frac{64}{45}\pi^3 \frac{\epsilon_c}{N^2}\frac{R^3}{\lambda^3}\omega,
\end{equation}
where $\epsilon_c=\frac{\epsilon-1}{3(\epsilon+2)}$ and $\epsilon$ is the relative dielectric constant of the nanoparticle.
 The above formula results in $\Lambda \approx 2.4 \times 10^{-4} \omega$ from photon recoil.
 
The surrounding gas pressure of $3\times10^{-9}$\,mbar results in a damping rate of $\gamma_\text{exp}=5.9\times10^{-5}$\,Hz at an environmental temperature of $60$\,K (thermal occupation $n_\text{th}=10^7$) resulting in a thermalisation rate $\gamma = n_\text{th} \gamma_\text{exp}\approx 2\times 10^{-3} \omega$. As a further reduction in pressure by 3 orders of magnitude seems feasible \cite{Tebbenjohanns_2021}, the thermalization rate can still be significantly suppressed below the localization rate as discussed in Appendix~\ref{apd:localisationdissipator}.

Accordingly, the effect of the environment is already reduced to the relevant range considered in Fig.~\ref{fig:fidvsgammalambda}.
Note, however, that the experimental parameter for the strength of the double-well Hamiltonian $H_\text{add}$ of ${\cal E}=8 \times 10^5 \hbar \omega$ significantly exceed the value chosen for the simulations of ${\cal E}=10 \hbar \omega$. On an experimental level, reducing the power sufficiently to match that constraint corresponds to performing the experiment at the level of $\omega \approx 1$\,Hz. While the normalized localization rate would not alter, a corresponding reduction of the pressure by 5 orders of magnitude would be required and also the feasibility of such optical traps is unclear. From a different perspective, this is however not necessary, as it may be expected that similar effects as discussed here will emerge also for large ${\cal E}$. Theoretical analysis of this behaviour however requires treatment of a significantly larger number of energy levels, which is hard to treat numerically.

\section{Conclusions}\label{sec:conclusions}
In this paper, we constructed an approach to study
the thermodynamics a nanoparticle under the combined action of a time-varying potential, which reproduce the transition from a harmonic to double-well landscape, and external dissipative influences. To simulate numerically its dynamics, we needed to discretise the system without loosing its dynamical features, a considerably complex task due to the presence of the non-trivial time-varying potential. Here, we provide a general approach to implement such a discretisation at a suitable level of approximation, and that can be also used to characterise the non-equilibrium thermodynamics of the system. The latter is here provided 
in terms of the Wehrl entropy, which well captures the non-Gaussian features of the system, and its production rates. The analysis showed that, starting from a thermal initial state relative to the initial potential, a non-equilibrium steady-state is eventually achieved. Moreover, after suitably setting the coupling strengths of the thermalisation and localisation processes, one can arrive at
a superposition state in position, namely a non-equilibrium high-coherence steady-state exhibiting two delocalised peaks in its position distribution. We demonstrated that this state survives the end of the variation of the potential and maintains its quantum features.
By suitably squeezing the initial thermal state, one can further improve the final coherence value.

Finally, we investigated the feasibility of implementing the proposed scheme in optical-levitation experiments. While state-of-the-art experiments operate at the relevant levels of decoherence, our study addresses a change in the trapping potential -- from quadratic to double-well -- involving low-energy states that is currently experimentally unexplored. 
It would be interesting to further our analysis 
to extend it to the context of electrostatic or magnetic levitation, and nanomechanical resonators (cf., for example, Ref.~\cite{Pistolesi_2021}), which might offer more favourable working regimes.

\acknowledgements

We acknowledge support from the Lise Meitner Programme: M2915, the Austrian Science Fund (FWF): Y 952-N36, START, the H2020-FETOPEN-2018-2020 project TEQ (grant nr. 766900), the DfE-SFI Investigator Programme (grant 15/IA/2864), the Royal Society Wolfson Research Fellowship (RSWF\textbackslash R3\textbackslash183013), the Leverhulme Trust Research Project Grant (grant nr.~RGP-2018-266), the UK EPSRC (grant nr.~EP/T028106/1).

\bibliography{references}

\appendix

\section{Derivation of irreversible entropy production rate for localisaiton}\label{apdsec:derivationofiep}

Here, we derive the irreversible entropy production rate  for the localisation in Eq.~\eqref{equ:localisationentropyproductionrate} starting from \eq{equ:localmastereqsc}. In the original derivation, which is given in Ref.~\cite{Santos_2018}, one starts from a different localisation dissipator with respect to the one in \eq{equ:localmastereqsc}. In particular, it reads
\begin{equation}
    D(\rho)=-\frac{\lambda}{2}[J_z,[J_z,\rho]],
\end{equation}
which has $J_x$ exchanged with $J_z$ when compared to \eq{equ:localmastereqsc}.

Our first step in the derivation of Eq.~\eqref{equ:localisationentropyproductionrate} is to rewrite \eq{equ:localmastereqsc} employing
the relation $J_x=\frac{1}{2}(J_++J_-)$, which gives
\begin{equation}\label{apdequ:localisation}
D_\text{lc}(\rho)=\frac{\Lambda}{4}\left\{ [J_-,f(\rho)] - [J_+, f^\dagger(\rho)] \right\},
\end{equation}
where we define the superoperator currents
\begin{equation}\label{apdequ:currents}
f(\rho) = - [J_+,\rho] - [J_-,\rho], \quad\text{and}\quad f^\dagger(\rho) = -f(\rho).
\end{equation}
The current $-f^\dagger(\rho)=f(\rho)=0$ when $\rho$ is localised.
Then, we follow Sec.~IV(B) in \cite{Santos_2018} where Takahashi- Shibata-Schwinger (TSS) approach was used to simplify the derivation \cite{Takahashi_1975}, although giving the same result as in the main text. 

The TSS approach maps the state $\ket{n_a}$ of the spin system into two bosonic modes $\ket{n_a,n_b}$ with the constraint $n_a+n_b=2j$. We refer to Ref.~\cite{Gyamfi_2019} for a detailed discussion on the TSS mapping. With two bosonic modes, one can define the bosonic coherent state as
\begin{equation}
        \ket{\alpha,\beta}=e^{-\frac{\mathcal{I}}{2}}\sum_{m=-j}^{j}\frac{\alpha^{j-m}}{\sqrt{(j-m)!}}\frac{\beta^{j+m}}{\sqrt{(j+m)!}}\ket{j-m,j+m}.
\end{equation}
Where we let $n_a=j-m, n_b=j+m$ with $m$ being the spin number, and denote the total amplitude as $\mathcal{I}=\lvert\alpha\rvert^2+\lvert\beta\rvert^2$. Therefore, given a general common density matrix $\rho$, the associated Husimi-Q function $\mQ(\alpha,\beta)=\frac{1}{\pi^2}\bra{\alpha,\beta}\rho\ket{\alpha,\beta}$ reads \cite{Scully_1994} 
\begin{equation}
    \mQ(\alpha,\beta)=\frac{e^{-\mathcal{I}}}{\pi^2} \mathcal{V}(\alpha,\beta),
\end{equation}
with $\mathcal{V}(\alpha,\beta)=\sum_{m,m'}\frac{\rho_{m,m'}(\alpha^\ast)^{j-m}(\beta^\ast)^{j+m}(\alpha)^{j-m'}(\beta)^{j+m'}}{\sqrt{(j-m)!(j+m)!(j-m')!(j+m')!}}$. Furthermore, define
\begin{equation}
    \alpha=\sqrt{\mathcal{I}}\cos{\frac{\theta}{2}}e^{-i\phi/2}, \quad \beta=\sqrt{\mathcal{I}}\sin{\frac{\theta}{2}}e^{i\phi/2},
\end{equation}
such that two sets of variables are related as
\begin{equation}\label{equ:scandtssrelation1}
    \D^2\alpha~\D^2\beta = \frac{\pi}{4}\mathcal{I}~\D\mathcal{I}~\D\Omega.
\end{equation}
In this way, one can interchange the TSS and the spin-coherent representations via the following relation \cite{Santos_2018,Vershynina_2012}
\begin{equation}
    \ket{\alpha,\beta}=\frac{e^{-\mathcal{I}/2}\sqrt{\mathcal{I}}^{2j}}{\sqrt{(2j)!}}\ket{\Omega}.
\end{equation}
Therefore one has that
\begin{equation}\label{equ:scandtssrelation2}
    \mQ(\alpha,\beta)=\frac{e^{-\mathcal{I}}\mathcal{I}^{2j}}{\pi^2(2j)!}\mQ(\Omega).
\end{equation}

In TSS representation, the derivative of Wehrl entropy reads
\begin{equation}\label{equ:wehrlentropyratetss}
\frac{\D S}{\D t}\rvert_\text{diss}=-\int\D^2\alpha\D^2\beta~\mD(\mQ)\ln\mQ,
\end{equation}
and the following correspondences hold
\begin{equation}
    \begin{split}
        [J_+,\rho]\rightarrow \mJ_+(\mQ)&=(\alpha^\ast\partial_{\beta^\ast}-\beta\partial\alpha)\mQ, \\
        [J_-,\rho]\rightarrow \mJ_-(\mQ)&=(\beta^\ast\partial_{\alpha^\ast}-\alpha\partial\beta)\mQ.
    \end{split}
\end{equation}
By applying them to \eq{apdequ:localisation}, one obtains
\begin{equation}\label{apdequ:localisationps}
\mD_\text{lc}(\mQ)=\frac{\Lambda}{4} \left\{ \mJ_-(f(\mQ)) - \mJ_+(f(\mQ)^\ast) \right\},
\end{equation}
with $f(\mQ)=\left( \alpha\partial_{\beta} - \alpha^\ast\partial_{\beta^\ast} + \beta\partial_{\alpha} - \beta^\ast\partial_{\alpha^\ast} \right) \mQ$.
By inserting the latter relation in \eq{equ:wehrlentropyratetss} and integrating by parts, we get
\begin{equation}
\frac{\D S}{\D t}\rvert_\text{diss}=\frac{\Lambda}{4} \int~\frac{\D^2\alpha\D^2\beta}{\mQ}~\left\{ f(\mQ)\mJ_-(\mQ) - f(\mQ)^\ast\mJ_+(\mQ) \right\}.
\end{equation}
We now use the relation $f(\mQ) = - f(\mQ)^\ast$ and $f(\mQ)^\ast=\mJ_-(\mQ)+\mJ_+(\mQ)$, and obtain
\begin{equation}\label{eq.S_diss}
\frac{\D S}{\D t}\rvert_\text{diss}
=\frac{\Lambda}{4} \int \frac{\D^2\alpha\D^2\beta}{\mQ}~ \lvert f(\mQ) \rvert^2 
\end{equation}
Such an equation is quadratic in terms of the current $f(\mQ)$ without linear terms. However, only linear terms provide the irreversible entropy production \cite{Santos_2018}. This means that the localisation dissipator is only associated to an entropy flux but not an entropy production. 
Finally, when expressing Eq.~\eqref{eq.S_diss} back in the spin-coherent representation with the help of Eqs.~\eqref{equ:scandtssrelation1} and \eqref{equ:scandtssrelation2}, we get \eq{equ:localisationentropyproductionrate}.

\section{The localisation dissipator}\label{apd:localisationdissipator}
In Lamb-Dicke regime ($k_0x_\text{zp}\ll 1$), the wave vector $k_0=2\pi/\lambda$ and zero-point motion $x_\text{zp}=\sqrt{\hbar/(2m\omega_m)}$). $m$ is the mass of the particle and $\omega_m$ is the trapping frequencies.

\paragraph{Dissipator due to gas pressure.}
The pressure of the gas results in a combination of two dissipators (need cites)
\begin{equation}
D_\text{gas} = D_\text b[\rho] + D_\text f[\rho],
\end{equation}
with each of them being
\begin{equation}
D_\text b[\rho] = - \frac{m\gamma k_\text{\tiny B} T}{\hbar^2} [x, [x,\rho]],
\end{equation}
\begin{equation}
D_\text f[\rho] = - i \frac{\gamma}{2\hbar} [x, \{p, \rho\}].
\end{equation}
T is the temperature of the chamber, $\gamma$ is the viscous friction that can be calculated from kinetic gas theory \cite{Chang1005}:
\begin{equation}
\gamma =\frac{64}{3} \frac{r^2P}{m v_\mathrm{gas}}.
\end{equation}
with the mean gas velocity $v_\mathrm{gas}=\sqrt{8k_bT/(\pi m_0)}$, and $m_0$ the atomic mass in kg.
If the protocol takes a couple of oscillation periods and the measurement time will show to be on the order of less than a ms, the term $D_\text f[\rho]$ does not alter the state and the measurement in a detectable way, which can be neglected. 

\paragraph{Dissipator due to photon recoil.}
Photon recoil is modelled by a position localization dissipator as well \cite{gardiner2004quantum}:
\begin{equation}
D_\text r[\rho] = - \Lambda'_\text r [x,[x,\rho]],
\end{equation}
with
\begin{equation}
\Lambda_r' = \frac{7\pi \epsilon_0}{30 \hbar} \left(\frac{\epsilon_c V E_0}{2\pi}\right)^2 k_0^5,
\end{equation}
where $V$ the volume of the particle. 
So the total dissipator takes the form 
\begin{equation}
D[\rho] = D_\text{gas}[\rho] + D_\text r[\rho] = -\Gamma' [x,[x,\rho]], 
\end{equation}
with $\Gamma' = \Gamma'_\text b + \Lambda'_\text r$.

\end{document}